\documentclass{mn2e}

\usepackage[dvips]{graphicx}
\usepackage{graphicx}
\usepackage{amsmath,amssymb}
\usepackage{epsf}
\usepackage{dcolumn}
\usepackage{color}
\usepackage{epsf}
\usepackage{psfrag}
\usepackage{epsfig}
\usepackage{bm}
\usepackage{threeparttable}
\usepackage{tensor}

\newcommand{\be}{\begin{equation}}
\newcommand{\ee}{\end{equation}}

\newcommand{\Om}{\Omega_m}

\newcommand{\OL}{\Omega_{\Lambda}}

\newcommand{\rco}{\rho_{c0}}

\newcommand{\rmr}{\rho_m}

\newcommand{\rL}{\rho_{\CC}}

\newcommand{\rLo}{\rho_{\CC 0}}

\newcommand{\CC}{\Lambda}

\newcommand\ringring[1]{%
  {
   \mathop{\kern0pt #1}\limits^{
     \vbox to-1.85ex{
       \kern-2ex 
       \hbox to 0pt{\hss\normalfont\kern.1em \r{}\kern-.45em \r{}\hss}%
       \vss 
     }
   }
  }
}

\begin{document}
\title[Density perturbations for running vacuum]{Density perturbations for running vacuum: a successful approach to structure formation and to the $\sigma_8$-tension}
\author[G\'omez-Valent and Sol\`a Peracaula]
{Adri\`a G\'omez-Valent\thanks{adriagova@fqa.ub.edu}, and Joan Sol\`a Peracaula\thanks{sola@fqa.ub.edu}
 \\
Departament  de F\'\i sica Qu\`antica i Astrof\'\i sica, and Institute of Cosmos Sciences, Universitat de Barcelona, Av. Diagonal 647\\
\vspace{0.1cm}
 E-08028 Barcelona, Catalonia, Spain
}

\date{\today}

\maketitle

\begin{abstract}
Recent studies suggest that dynamical dark energy (DDE) provides a better fit to the rising affluence of modern cosmological observations than the concordance model ($\CC$CDM) with a rigid cosmological constant, $\CC$. Such is the case with the running vacuum models (RVMs) and to some extent also with a simple XCDM parametrization.
Apart from the cosmic microwave background (CMB) anisotropies, the most crucial datasets  potentially carrying the DDE signature are: i) baryonic acoustic oscillations (BAO), and ii) direct large scale structure (LSS) formation data (i.e. the observations on $f(z)\sigma_8(z)$ at different redshifts). As it turns out, analyses mainly focusing on CMB and with insufficient BAO+LSS input, or those just making use of gravitational weak-lensing data for the description of structure formation, generally fail to capture the DDE signature, whereas the few existing studies using a rich set of CMB+BAO+LSS data (see in particular  Sol\`a, G\'omez-Valent \& de Cruz P\'erez 2015, 2017; and  Zhao et al. 2017) do converge to the remarkable conclusion that DDE might well be encoded in the current cosmological observations. Being the issue so pressing, here we explain both analytically and numerically the origin of the possible hints of DDE in the context of  RVMs, which arise at a significance level of $3-4\sigma$.  By performing a detailed study on the matter and vacuum perturbations within the RVMs, and comparing with the XCDM, we show why the running vacuum fully relaxes the existing $\sigma_8$-tension and accounts for the LSS  formation data much better than the concordance model.
\end{abstract}



\section{Introduction}

The positive acceleration of our universe was discovered almost twenty years ago thanks to the accurate measurement of the apparent magnitude versus redshift relation of distant Supernovae of type Ia (SNIa) (Riess et al. 1998; Perlmutter et al. 1999). Assuming that General Relativity is the correct classical theory of gravity (also at cosmological scales), one needs to introduce a new component into the Cold Dark Matter (CDM) model to generate the negative pressure that is needed to explain such late-time speeding-up.  The generic name for it is dark energy (DE). After  two decades of intense research, the exact nature of the DE  remains unknown, but observations tell us that it must fulfill two basic requirements: (i) the current associated equation of state (EoS) for the DE  must be of vacuum or quasi-vacuum type, i.e. $w\simeq -1$; and (ii) its clustering properties, if available,  must be very much suppressed at deep subhorizon scales, meaning that the DE must be essentially uniform and hence evenly distributed in all corners of the universe.

The ``simplest'' proposed  framework satisfying these conditions is obtained by adding a tiny cosmological constant (CC) in Einstein's field equations, $\CC>0$. This setting warrants a universe with a uniformly distributed vacuum energy density  $\rLo=\Lambda/(8\pi G)=$const. ($G$ being Newton's constant)  with EoS parameter  exactly $w=-1$.  Such configuration of  vacuum energy is automatically unable to cluster. The resulting theoretical construction, usually assumed spatially flat in order to be consistent with an early period of inflation, is the so-called concordance or $\Lambda$CDM model (Peebles 1984, 1993). It  is considered the standard model of cosmology and is able to explain with proficiency a wide variety of cosmological observations, including of course the high precision Cosmic Microwave Background (CMB) data (cf. Planck Collab. XIII 2016; DES Collab. 2017). But despite its numerous successes, the CC is also at the root of two of the most profound theoretical problems in physics, namely the old CC problem (Weinberg 1989; Padmanabhan 2003; Sol\`a 2013) and the Cosmic Coincidence problem (see e.g. the reviews by  Peebles \& Ratra 2003; Copeland, Sami \& Tsujikawa 2006), both of them  still lacking a solution. In addition, there exist some severe and persistent tensions between data sets in the context of the $\Lambda$CDM model. They involve relevant parameters of cosmology, such as the Hubble parameter, i.e. the current value of the Hubble function, $H(t_0)=H_0$, and the current value of the rms of mass fluctuations at spheres of $8\,h^{-1}$ Mpc, i.e. the parameter $\sigma_8$,  where  $h=H_0/(100\,{\rm km/s/Mpc})$  stands for the reduced Hubble parameter.

Among the many alternative scenarios beyond the $\Lambda$-term proposed throughout the years to solve these conundrums one finds a large body of DE entities, including quintessence and the like, see e.g. the comprehensive book (Amendola \& Tsujikawa 2010) and references therein. Not all the models perform equally good, though. Previous works in the literature have shown that some dynamical DE models are  able to fit considerably better the cosmological data than the standard $\Lambda$CDM with a rigid $\CC$-term. The positive signal of DE dynamics can be captured in different ways and confidence levels, to wit:  i) using a simple XCDM parametrization; ii) a  nontrivial $\phi$CDM scalar field model with a specific potential (see e.g. Sol\`a, G\'omez-Valent \& de Cruz P\'erez 2017a,b); iii) a non-parametric reconstruction of the DE equation of state as a function of the redshift (Zhao et al. 2017); iv) and also with a variety of dynamical vacuum models (DVMs), more conspicuously those in the class of the so-called running vacuum models (RVMs) (see Sol\`a, G\'omez-Valent \& de Cruz P\'erez 2015, 2017a,b,c,d; and also Sol\`a, de Cruz P\'erez \& G\'omez-Valent 2018).

In the context of the RVMs, the vacuum energy density evolves (runs) slowly with the cosmic expansion. The law describing its evolution is motivated from the renormalization group formalism of Quantum Field Theory (QFT) in curved spacetime (for reviews see  Sol\`a  2011, 2013,  2016;  Sol\`a \& G\'omez-Valent 2015, G\'omez-Valent 2017, and references therein).  After the inflationary epoch (whose evolution can also be described in this context, see e.g. Lima, Basilakos  \& Sol\`a  2013, 2015; Sol\`a  2015), the vacuum density takes on the following simple form, $\rho_\Lambda(H)=C_0+C_1 H^2$.  In such framework it should be possible to better tackle the basic CC problems. Actually, the RVMs are not only well motivated from a theoretical point of view, but are also preferred over the $\Lambda$CDM at an outstanding $\sim 4\sigma$ c.l. when they are confronted to the same string of rich enough cosmological data sets SNIa+BAO+$H(z)$+LSS+CMB, which includes not only the data on SNIa and Hubble function at different redshifts, but also the crucial information encoded in the CMB anisotropies, the data on Baryon Acoustic Oscillations (BAOs) and the Large Scale Structure (LSS), see the above mentioned papers. The last three data sources have proved to be indispensable to detect the signature of vacuum dynamics.

In particular, it is of utmost importance to incorporate the LSS data from the weighted growth rate ($f(z)\sigma_8(z)$)  provided by different galaxy surveys, as e.g. BOSS (Gil-Mar\'in et al. 2017), which are mainly (but not only) extracted from the analysis of Redshift Space Distortions (RSD). As shown in extensive numerical analyses (Sol\`a, G\'omez-Valent \& de Cruz P\'erez 2015, 2017a,b,c,d;  Sol\`a, de Cruz P\'erez \& G\'omez-Valent 2018) the RVMs are capable of substantially improving the fit of the LSS observations while keeping the quality fit to the BAO+CMB part. This is mainly due to the fact that the RVMs allow a $8-9\%$ reduction in the value of the $\sigma_8$ parameter with respect to the typical values that are obtained in the $\Lambda$CDM, and this loosens the tension with the data obtained from RSD (see e.g. Macaulay, Wehus \& Eriksen 2013;  Basilakos \& Nesseris 2016, 2017) and from weak gravitational lensing (see e.g. Heymans et al. 2013; Hildebrandt et al. 2017; Joudaki et al. 2018). For a devoted study of the impact of the RVMs to the issue of the $H_0$ and $\sigma_8$ tensions, see  Sol\`a, G\'omez-Valent \& de Cruz P\'erez 2017d. See also Valentino et al. 2016, 2017 for related studies.

Due to the crucial role played by the LSS data in the overall fit, it is extremely important to compute the linear perturbations correctly in order to ensure the correct inference of the model parameters and the right determination of the significance of the detected signal. In actual fact, the lack of systematically taking into account the LSS data in the overall fit to the cosmological data is at the root of missing the possible dynamical DE effects in most past studies in the literature, including Planck Collab. XIII 2016 and DES Collab. 2017. To the best of our knowledge, the first studies duly taking into account these effects are those by Sol\`a, G\'omez-Valent \& de Cruz P\'erez, 2015, 2017a,b; Sol\`a, de Cruz P\'erez \& G\'omez-Valent 2018; and  Zhao et al. 2017. They resonate  in the important conclusion of favoring dynamical DE at a $3-4\sigma$ c.l.

The main aim of this work is to study in detail the linear density perturbations in the RVMs with a vacuum-matter interaction and provide an analytical explanation for the origin of these important dynamical DE effects in such context, after the explicit numerical analysis has already supported such level of evidence. We want to illustrate how the RVMs seem to be an ideal framework to describe the LSS data and relax the aforementioned $\sigma_8$-tension with the $\CC$CDM prediction. The RVMs indeed provide a possible natural solution to the $\sigma_8$-tension, as advanced in  (G\'omez-Valent \& Sol\`a 2018). At the same time we wish to confront our results for the RVMs with those obtained from the XCDM as a baseline model for comparison used in generic studies of DDE.  The XCDM  (or $w$CDM) is the next-to-simplest extension of the $\CC$CDM  and is characterized by the EoS  $p=w\rho$, with $w=$const., in which $w=-1$ corresponds to the $\CC$CDM (Turner \& White 1997).  One expects that if a particular model is capable of detecting significant traces of DDE in the data should also be corroborated by some departure of $w$ from $-1$ when the same data are analyzed in terms of the XCDM parametrization.

The layout of this paper is as follows. In Sect. 2 we derive the equations that govern the evolution of density perturbations at subhorizon scales during the matter and vacuum-dominated epochs for general DVMs using the Newtonian conformal gauge. In Sect. 3 we define the canonical RVM in interaction with matter. The relative size of vacuum energy density fluctuations in this model are analyzed in Sect. 4. We reconsider the situation in the synchronous gauge in Sect. 5.  The connection between the weighted linear growth $f(z)\sigma_8(z)$ and matter power spectrum for the RVM is outlined in Sect. 6, whereas Sect. 7 is devoted to the leading effects of running vacuum on that LSS observable. Finally, Sect. 8 studies the implications on the weak-lensing parameter  $S_8$.  In Sect. 9 we deliver our conclusions.


\section{Density perturbations with vacuum dynamics at subhorizon scales in the Newtonian gauge}\label{sect:NewtonianGauge}

In what follows we discuss the cosmological density perturbations for the spatially flat Friedmann-Lema\^itre-Robertson-Walker (FLRW) metric  in the context of the dynamical vacuum models (DVMs), which have been recently discussed in the literature from different points of view (see e.g. Sol\`a, G\'omez-Valent \& de Cruz P\'erez 2015, 2017a,b,c,d).  For these models  the vacuum energy density $\rL$  is not constant  but dynamical, meaning that the EoS parameter is still $w=-1$ but the corresponding vacuum energy density evolves with the expansion. The evolution of $\rL$ is sufficiently small as to depart  mildly at present  from the rigid assumption $\rL=$const. of the $\CC$CDM. For the DVMs we may assume that $\rL=\rL(\zeta)$, where $\zeta(t)$ is a cosmic variable, typically it can be the  scale factor or even the Hubbe function $H$. The considerations made in the present section will be general for any DVM, but from Section 3 onwards we shall consider a specific type of DVM in which $\zeta=H$, called the running vacuum models (RVMs). Although there are several possible realizations of the RVMs we will focus here in the canonical type, which will be introduced in Sect. 3.

In the conformal Newtonian gauge (or longitudinal gauge) we write the perturbed FLRW metric in conformal time $\eta$ as follows (Mukhanov, Feldman \&  Brandenberger 1992):
\begin{equation}
ds^2=a^2 (\eta)\left[(1+2\Phi)d\eta^2-(1+2\Psi)d\vec{x}^2\right]\,,
\end{equation}
where $\Phi$ and $\Psi$ are the so-called Bardeen potentials (Bardeen 1980), and we recall that $d\eta=dt/a$, with $t$ the cosmic time. Treating the matter and vacuum components as perfect fluids, it can be shown that the
\begin{table*}
\begin{center}
\begin{scriptsize}
\resizebox{1\textwidth}{!}{
\begin{tabular}{ |c|c|c|c|c|c|c|c|c|}
\multicolumn{1}{c}{Model} &  \multicolumn{1}{c}{$H_0$(km/s/Mpc)} &  \multicolumn{1}{c}{$\omega_b$} & \multicolumn{1}{c}{{\small$n_s$}}  &  \multicolumn{1}{c}{$\Omega_m$} &\multicolumn{1}{c}{$\nu$} &\multicolumn{1}{c}{$w$} &\multicolumn{1}{c}{$\ln A$} &\multicolumn{1}{c}{$\ln B$} \vspace{0.5mm}
\\\hline
$\Lambda$CDM  & $68.83\pm 0.34$ & $0.02243\pm 0.00013$ &$0.973\pm 0.004$& $0.298\pm 0.004$ & - & -1 & - & - \\
\hline
XCDM  & $67.16\pm 0.67$& $0.02251\pm0.00013 $&$0.975\pm0.004$& $0.311\pm0.006$ & - &$-0.936\pm{0.023}$ & 2.68 & 1.56 \\
\hline
RVM  & $67.45\pm 0.48$& $0.02224\pm0.00014 $&$0.964\pm0.004$& $0.304\pm0.005$ &$0.00158\pm 0.00041 $ & -1 & 6.74  & 5.62 \\
\hline
\end{tabular}}
\end{scriptsize}
\caption{Best-fit values obtained from the SNIa+BAO+$H(z)$+LSS+CMB fitting analysis of (Sol\`a, G\'omez-Valent \& de Cruz P\'erez 2017d) for the $\CC$CDM, XCDM and the RVM, together with the Akaike and Bayesian
evidence criteria. See the aforementioned paper for further details, including the complete list of data used in the analysis and the corresponding references. Both, the XCDM and, more conspicuously, the RVM, are clearly preferred over the $\Lambda$CDM. The positive signal in favor of vacuum dynamics reaches $\sim 3.8\sigma$ c.l. in the RVM, whereas in the XCDM parametrization the signal of DE dynamics is lower ($\sim 2.8\sigma$ c.l.), but still significant.}
\end{center}
\label{tableFit1}
\end{table*}
 fulfillment of the ij component of the perturbed Einstein's equations, i.e. $\delta G_{ij}=8\pi G\delta T_{ij}$, requires $\Psi=-\Phi$, and this relation will be assumed from now on. In the presence of anisotropic stress such relation would not hold, for instance.  Let us also note that, in this gauge, the vector and the tensor degrees of freedom are eliminated from the beginning. In fact, the vector part of the perturbation is set to zero  and the non-diagonal spatial part decouples from the rest in the form of gravitational waves propagating in the FLRW background.

 The equations that govern the growth of the perturbations in this gauge can be found by the standard procedure (see e.g. Ma \& Bertschinger 1995). As independent perturbations equations we can take the $(00)$ component of  the perturbed Poisson equation,
\begin{equation}\label{eq:nablaequation}
 3\mathcal{H}^2\Phi-\Delta\Phi+3\mathcal{H}\dot{\Phi} = -4\pi G a^2\sum_{i=\Lambda,m}\delta\rho_i\,,
\end{equation}
the perturbed local energy-momentum conservation equation  $\nabla^{\mu}T_{\mu 0}=0$,
\begin{equation}\label{eq:ContinuityOriginal}
\sum_{i=\Lambda,m} \dot{\delta\rho_i}+(p_i+\rho_i)(\Delta v_i-3\dot{\Phi})+3\mathcal{H}(\delta\rho_i+\delta p_i)=0\,,
\end{equation}
and of its spatial part $\nabla^{\mu}T_{\mu i}=0$, which leads to the (perturbed) Euler equation
\begin{equation}\label{eq:Euler1}
\sum_{i=\Lambda,m} \frac{d}{d\eta}\left[(p_i+\rho_i)v_i\right]+(\rho_i+p_i)(4\mathcal{H}v_i+\Phi)+\delta p_i=0\,.
\end{equation}
Throughout the paper an overdot denotes a derivative with respect to the conformal time, $\dot{f}=df/d\eta$, $\mathcal{H}=\dot{a}/a$ is the Hubble function in conformal time, and $\Delta$ is the Laplace operator with respect to the comoving coordinates. Differentiation with respect to the cosmic time and the scale factor will also be used and a different notation will be employed.
Furthermore, in the above equations it is understood  that $v_i$ stands for the (longitudinal) velocity potential. In fact, the longitudinal contribution of the (peculiar) 3-velocity of the i$_{th}$ component can be written in terms of the gradient of the scalar velocity potential, specifically  $\vec{v}_i^L=\vec{\nabla} v_i$ with $i=\Lambda,m$. Recall that the transverse part of the 3-velocity only affects the vector perturbations, which are decoupled from the scalar ones and are not being considered here (see e.g. Gorbunov \& Rubakov 2011 for further details) \footnote{For this reason we shall henceforth suppress the upper index $L$ in $\vec{v}_i^L$ for each component.}. At physical scales $\lambda$ deeply inside the horizon, i.e. $\lambda\ll 3000\,h^{-1}{\rm Mpc}$,  and taking into account that in the DVMs under consideration the vacuum interacts with the matter sector, the above equations can be written in momentum space as follows:
\begin{align}
k^2\Phi&=-4\pi G a^2(\delta\rho_m+\delta\rho_\Lambda)\,,\label{eq:PoissonNewton}\\
      0&=\dot{\delta\rho_m}+3\mathcal{H}\delta\rho_m-k^2v_m\rho_m+\dot{\delta\rho_\Lambda}\,,\label{eq:ContinuityNewton}\\
       0&=\frac{d}{d\eta}\left(\rho_mv_m\right)+4\mathcal{H}\rho_mv_m+\rho_m\Phi-\delta\rho_\Lambda\,,\label{eq:EulerNewton}
\end{align}
where $k\equiv|\vec{k}|$ is the comoving wave number (hence $k/a$ is the physical one), and $\rho_m$ is the sum of the mean baryon and dark matter energy densities in the universe, i.e. $\rho_m=\rho_b+\rho_{dm}$. Since we are mainly interested in the physics at subhorizon scales ($k^2\gg\mathcal{H}^2$)  we have dropped the terms that are suppressed by this condition, e.g. when going
 from Eq.\,\eqref{eq:nablaequation} to Eq.\,\eqref{eq:PoissonNewton}. We proceed systematically with this approximation throughout our exposition.
In the previous equations,
\begin{equation}\label{eq:vm}
  v_m=\frac{v_{dm}\rho_{dm}+v_b\rho_b}{\rho_m}
\end{equation}
is the total matter velocity potential, obtained upon weighting the contributions of the two matter components. We are interested in the total matter growth because this is usually what the LSS observables are sensitive to. For instance, the RSD are caused by the total amount of matter, not only by one particular type. Notice also that we are studying the evolution of non-relativistic matter and vacuum perturbations from the deeply matter-dominated (MD) epoch up to the present time. In this period of the cosmic expansion the radiation energy density only constitutes a derisory fraction of the critical energy density in the universe, and moreover it is completely decoupled from matter, so we can neglect the effect of radiation at both, background and perturbations levels.

Using the background continuity equation for a general DVM in interaction with matter,
\begin{equation}\label{eq:contBackground}
\dot{\rho}_\Lambda+\dot{\rho}_m+3\mathcal{H}\rho_m=0\,,
\end{equation}
one can write \eqref{eq:EulerNewton} in a more standard way:
\begin{equation}\label{eq:EulerNewton2}
\dot{v}_m+\mathcal{H}v_m+\Phi+\psi v_m-\frac{\delta\rho_\Lambda}{\rho_m}= 0\,,
\end{equation}
with $\psi\equiv-\dot{\rho}_\Lambda/\rho_m$ (not to be confused with the potential $\Psi$, which was previously fixed as $\Psi=-\Phi$ once and for all in this study). The first three terms of the last expression correspond to those appearing in the (perturbed) Euler equation within the $\Lambda$CDM. In fact, they can be obtained in a simple way, just by perturbing the Newtonian gravitational law,
\begin{equation}
\frac{d\vec{v}_{p}}{dt}=\vec{a}_{\rm cosm}-\frac{1}{a}\vec{\nabla}\Phi\,,\ \ \ \ \ \vec{v}_p=\frac{d\vec{r}}{d t}=\mathcal{H}\vec{x}+\vec{v}_m\,,
\end{equation}
where $\vec{v}_p$ is  the perturbed 3-velocity in proper coordinates and $\vec{a}_{\rm cosm}=\dot{\mathcal{H}}\vec{x}/a$ is the cosmic acceleration associated to the uniform-expansion observers. Recall that  $\vec{v}_m=\vec{\nabla} v_m$ and that $\vec{\nabla}$ denotes the gradient with respect to the comoving coordinates, and hence $(1/a)\vec{\nabla}$ is the gradient with respect tot the physical coordinates. Using $dt=a d\eta$ the above equation can be written as
\begin{equation}
\vec{\nabla}\left(\dot{v}_m+\mathcal{H}v_m+\Phi\right)=\vec{0}\,,
\end{equation}
which  indeed  leads to the standard Euler equation in the concordance model, if the integration constant is set to zero.
 The last two  terms of Eq. \eqref{eq:EulerNewton2} are new and can be interpreted as the change in the matter velocity potential that is induced by the matter-vacuum interaction. We should ask ourselves now if it is actually possible that this interaction modifies somehow the velocity of the matter particles. The loss of energy of the vacuum sector can only occur in two different ways: by vacuum decay through the generation of particle pairs, or because of an increase of the particles' masses. If the vacuum decay occurs only due to an increase of the particles' masses and we assume that the equivalence principle is preserved, we expect to recover the standard Euler equation (Koyama, Maartens \& Song  2009). This reasoning leads us to impose an extra relation in order to ensure the correct physical behavior of the DVMs at the linear perturbations level,
\begin{equation}\label{eq:ExtraRelation}
\delta\rho_\Lambda=\psi v_m\rho_m\,,
\end{equation}
so that the usual ($\Lambda$CDM) Euler equation is warranted:
\begin{equation}\label{eq:EulerNewton3}
\dot{v}_m+\mathcal{H}v_m+\Phi=0\,.
\end{equation}
It is also interesting to note that Eq. \eqref{eq:ExtraRelation} can also be written in terms of the physical velocity of matter and the gradient of vacuum perturbations,
\begin{equation}
\vec{\nabla}\delta\rho_\Lambda=-\dot{\rho}_\Lambda\vec{v}_m\,.
\end{equation}
In contrast, if particles pop out from the vacuum, then one does not expect {\it a priori}  an exact fulfillment of the Euler equation \eqref{eq:EulerNewton3}. In order to study this alternative scenario it is convenient to split \eqref{eq:EulerNewton2} as follows,
\begin{align}
\dot{v}_m+\mathcal{H}v_m+\Phi+\psi v_m &= \mathcal{B}\,,\label{eq:split1}\\
      \frac{\delta\rho_\Lambda}{\rho_m}&=\mathcal{B}\,,\label{eq:split2}
\end{align}
where $\mathcal{B}$ must be a linear function of the perturbed quantities under consideration. In addition, it is obvious that $\mathcal{B}$ must be proportional to the background function $\psi$ because if the vacuum energy density remains constant, i.e. if $\psi=0$, we must retrieve the $\Lambda$CDM result. This means that equations \eqref{eq:split1} and \eqref{eq:split2} must decouple from each other in this case, i.e. $\mathcal{B}=0$, so the vacuum perturbations disappear and the Euler equation is recovered. Thus, we expect $\mathcal{B}$ to take the following general form,
\begin{equation}
\mathcal{B}=\psi\sum_{i}\alpha_i\delta A_i\,,
\end{equation}
$\alpha_i$ being dimensionless constants and $\delta A_i$ perturbed functions with dimensions of inverse of energy. Notice that the choice $\mathcal{B}=\psi v_m$ satisfies the above mentioned conditions. This is precisely the relation that is obtained from \eqref{eq:ExtraRelation}, which is exactly fulfilled when the vacuum loses energy due to an increase of the matter particles' masses. From now on we will adopt \eqref{eq:ExtraRelation} for simplicity. This assumption allows us to study the two possibilities of vacuum decay with the same formula, although it is important to keep in mind that more general expressions for $\mathcal{B}$ could in principle apply if the vacuum decay occurred through particle creation.

To sum up, by using the Newtonian conformal gauge and applying the above arguments, the system of equations that governs the linear density perturbations at deep subhorizon scales in the DVMs are: \eqref{eq:PoissonNewton}, \eqref{eq:ContinuityNewton}, \eqref{eq:ExtraRelation}, and \eqref{eq:EulerNewton3}.

\begin{figure*}
\includegraphics[scale=0.55]{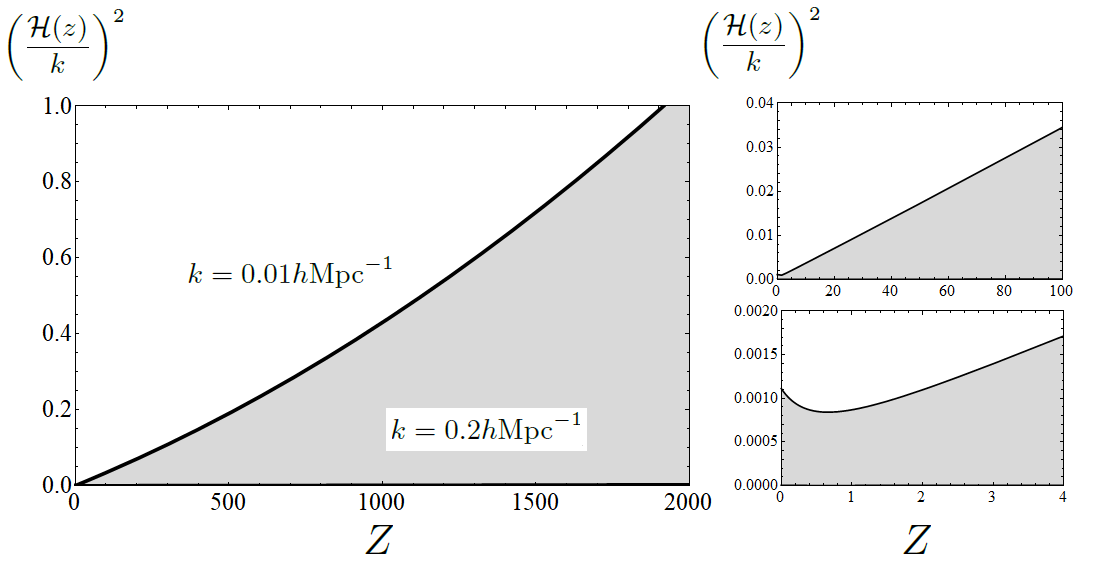}
\caption{{\it Left plot:} Squared ratio between the comoving scales $1/k$ and the comoving Hubble horizon, $\mathcal{H}^{-1}=H^{-1}/a$, as a function of the redshift.  The used range of comoving wave numbers, $k$,  correspond to those that  we have observational access to inside the horizon and at which the linear perturbations regime is still valid, namely the modes  $0.01\,h{\rm Mpc}^{-1}\leq k\leq 0.2\,h{\rm Mpc}^{-1}$. They are inside the gray band. The lowest modes (corresponding to the largest scales) reentered the horizon far in the past, at $z\simeq 1920$, previously to the decoupling time but already during the MD epoch, whereas the largest modes (smallest scales) reentered the horizon deeply in the radiation-dominated era, at $z\simeq 61.1\times 10^{3}$ (hence out of the plot); {\it Upper-right plot:} As in the left plot, but here for a shorter redshift range, up to $z=100$. This is to show that the modes we are focusing our attention on, i.e. those that lie in the gray region, are deeply inside the horizon from $z\sim 100$ up to the present time, i.e. $(\mathcal{H}/k)^2<0.035\ll 1$. This is all the more true at $ z\lesssim 10$  since the relevant modes then satisfy $(\mathcal{H}/k)^2<0.004\ll 1$. This legitimates to use the subhorizon approximation at these scales, see the text for further details; {\it Lower-right plot:}  Similar to the previous cases,  now in the narrow range $0\leq z\leq 4$, allowing us to appreciate the existence of a minimum at $z_{{\rm min}}\sim 0.6-0.7$. The latter  indicates the transition from a decelerated to an accelerated universe, which causes that the  lowest subhorizon modes start exiting the Hubble horizon again. The transition point,  defined as the point at which the deceleration parameter vanishes, i.e. $q(a_t)=-1-\dot{H}(a_t)/[a_tH^{2}(a_t)]=0$, can be analytically computed in the RVM: $a_t=\left[\frac{(1-3\nu)\,\Om}{2(\OL-\nu)}\right]^{1/(3(1-\nu))}$. Using the values of the RVM parameters in Table 1 we obtain $z_{t}=a_t^{-1}-1=0.663$.
\label{fig:aHoverk}}
\end{figure*}

\section{Running vacuum in interaction with matter}

Running vacuum models are particularly motivated realizations of the DVMs discussed above, in which the dynamical origin of the vacuum energy density can be conceived  from the point of view of the renormalization group (see e.g. Sol\`a 2011,  2013, 2015, 2016;  Sol\`a \& G\'omez-Valent 2015; G\'omez-Valent 2017, and references therein).  Hereafter we shall focus on the canonical or simplest form, and we will call it the RVM (running vacuum model).  In this case,  $\rL$ evolves with the Hubble rate: $\rL=\rL(H)$.  In this context one can say that  $\rL$  ``runs'' with the cosmic expansion.

The RVM  has a smooth $\CC$CDM limit, namely it departs  from the usual $\rL=$const. assumption characteristic of the $\CC$CDM through a continuous parameter $\nu$.  For $\nu=0$ the concordance model is recovered.  Specifically,  $\rL$ takes on the form
\begin{equation}\label{eq:RVMvacuumdadensity}
\rho_\CC(H) = \frac{3}{8\pi{G}}\left(c_{0} + \nu{H^2}\right)\,.
\end{equation}
%
Here $c_0=H_0^2\left(1-\Omega_m-\nu\right)$ is fixed by the boundary condition $\rL(H_0)=\rho_{\Lambda 0}=\rco\,(1-\Omega_m)$, with $\Omega_m=\Omega_b+\Omega_{dm}$ the nonrelativistic matter density parameter at present and $\rco$ the current critical density. The dimensionless coefficient $\nu$ is the vacuum parameter of the RVM. A nonzero value of it makes possible the cosmic evolution of the vacuum. It is expected to be very small, $|\nu|\ll1$, since the model must remain sufficiently close to the $\CC$CDM.  The moderate dynamical evolution of $\rL(H)$ is possible thanks to the vacuum-matter interaction, see Eq.\,(\ref{eq:contBackground}). Formally, $\nu$ can be given a QFT meaning by linking it to the $\beta$-function of the running $\rL$ (Sol\`a 2013; Sol\`a \& G\'omez-Valent 2015, and references therein). Theoretical estimates place its value in the ballpark of $\nu\sim 10^{-3}$ at most in the context of a typical Grand Unified Theory (GUT) (Sol\`a 2008), and this is precisely the order of magnitude for  $\nu$  preferred by the cosmological data (cf. G\'omez-Valent \& Sol\`a 2015; G\'omez-Valent, Sol\`a \& Basilakos 2015; Sol\`a, G\'omez-Valent \& de Cruz P\'erez 2015, 2017a,b,c,d; Sol\`a, de Cruz P\'erez \& G\'omez-Valent 2018). The order of magnitude coincidence between theoretical expectations and phenomenological fits to the data is quite reassuring.
Different realizations of the RVM are possible, but here we limit ourselves to the model studied in (Sol\`a, G\'omez-Valent \& de Cruz P\'erez 2017b,c,d; Sol\`a, de Cruz P\'erez \& G\'omez-Valent 2018), in which the vacuum exchanges energy only with dark matter.  Baryons and radiation are covariantly conserved and, therefore, they obey the same dilution laws under expansion as in the $\Lambda$CDM \footnote{See Appendix A for the treatment of the matter perturbations under the condition of baryon conservation.}:
\begin{equation}
\rho_b(a)=\rho_{b0}a^{-3}\qquad \qquad\rho_r(a)=\rho_{r0}a^{-4}\,.
\end{equation}
The corresponding normalized Hubble rate $E\equiv H/H_0$ (with $H=\mathcal{H}/a$) is
\begin{eqnarray}\label{eq:H2RVM}
E^2(a) &=& 1 + \frac{\Omega_m}{1-\nu}\left(a^{-3(1-\nu)}-1\right) \label{HRVM}\\ \nonumber\\
&& + \frac{\Omega_r}{1-\nu}\left(\frac{1-\nu}{1+3\nu}a^{-4} + \frac{4\nu}{1+3\nu}a^{-3(1-\nu)} -1\right)\,,\nonumber
\end{eqnarray}
and the total matter energy density reads
\begin{equation}\label{eq:rhoM}
\rho_{m}(a) =\rho_{m0}\,a^{-3(1-\nu)}+\frac{4\nu\rho_{r0}}{1 + 3\nu}\,\left(a^{-3(1-\nu)} - a^{-4}\right)\,.
\end{equation}
Note that for $\nu=0$ we recover the $\CC$CDM expressions, as it should be expected.

The numerical values of the parameters for the RVM used in all the calculations and plots of the present work have been obtained from the fitting analysis of (Sol\`a, G\'omez-Valent \& de Cruz P\'erez 2017d) based on a large string of SNIa+BAO+$H(z)$+LSS+CMB data described there. They are written in Table 1 for convenience, together with the values obtained in the same analysis for the $\Lambda$CDM model and the XCDM parametrization. Recall that  the XCDM (or $w$CDM) is the next-to-leading formulation of the DE beyond the $\CC$CDM. Rather than assuming a rigid CC-term $\CC=$const.  with exact EoS $w=-1$, one assumes that the DE obeys $p=w\rho$, with constant $w=-1+\epsilon$, such that for $w=-1$ (i.e. $\epsilon=0$) one retrieves the particular $\CC$CDM model.
It is natural to expect that if there are significant traces of DDE in the current observational data it should be possible to minimally capture them in a model-independent way through the XCDM parametrization by finding a small departure of $w$ from $-1$. Small departures of the EoS parameter from $-1$, i.e.  $|\epsilon|\ll1$,  would point to dynamical DE of quintessence ($w>-1$, i.e. $\epsilon>0$) or phantom ($w<-1$, i.e. $\epsilon<0$) type. Let us note that despite in the RVM the EoS is the strict vacuum one ($w=-1$), such vacuum energy density is dynamical. As a result, from Eq.\,(\ref{eq:RVMvacuumdadensity}) it is clear that if $\nu>0$ the vacuum energy density is larger in the past than is at present, and hence the RVM effectively behaves as quintessence. In contrast, if $\nu<0$ the effective behavior of the RVM would be phantom-like. From the best-fit values that we have found in Table 1, which take into consideration the indicated large set of cosmological data, we infer that the actual behavior of the RVM is quintessence-like at $3.8\sigma$ c.l., namely  $\nu=0.00158\pm 0.00041$. We can see from Table 1 that this is consistent with the EoS value that we have found for the XCDM, which is $w=-0.936\pm 0.023$  and hence favoring quintessence at about $2.8\sigma$ c.l.  The two signals are clearly pointing to the same direction,  but the RVM seems to involve a stronger germ of DDE than the simple XCDM parametrization.

We have carried out several tests in order to study the robustness of our results. Here we report on the outcome after performing an update of our database with respect to the one used in (Sol\`a, G\'omez-Valent \& de Cruz P\'erez 2017d). The list of changes is the following: we have added the data point $H(z=0.47)$ obtained with the differential-age technique by Ratsimbazafy et al. (2017), the anisotropic BAO and LSS data at $z_{\rm eff}=1.52$ from (Gil-Mar\'in et al. 2018), the weak-lensing data from (Hildebrandt et al. 2017), the LSS data point at $z=1.36$ reported in (Okamura et al. 2016); we have also replaced the SDSS LSS point from (Feix, Nusser \& Branchini 2015) by the one from (Shi et al. 2017), the 2MTF LSS point from (Springob et al. 2016) by the one from (Howlett et al. 2017), the LSS data point from (Granett et al. 2015) by those from (Pezzota et al. 2017), the BAO data at $z=0.106$ from the 6dFGS (Beutler et al. 2011) and at $z=0.15$ from SDSS DR7 (Ross et al. 2015) by their combined value at $z_{\rm eff}=0.122$ (Carter et al. 2018), the BAO Ly$\alpha$ forest data from (Delubac et al. 2015; Aubourg et al. 2015) by those from (du Mas des Bourboux et al. 2017), and the information of the 740 SNIa of the joint light-curve analysis (JLA) sample (Betoule et al. 2014) by the 15 SNIa from the CANDELS and CLASH Multy-Cycle Treasury programs obtained by the Hubble Space Telescope (Riess et al. 2018a), together with the 1049 SNIa of the Pantheon compilation (Scolnic et al. 2017), which also incorporates those from the JLA sample. Use has been made of the compressed version of these SNIa data provided in (Riess et al. 2018a). The impact of all these changes on the value of the RVM parameter is not dramatic at all:  the result of the new fit reads  $\nu=0.00134\pm 0.00038$,  still keeping a remarkable $3.53\sigma$ departure with respect to the standard $\Lambda$CDM ($\nu=0$). Furthermore, the obtained value of $\nu$ and of all the other fitted parameters are fully compatible with the ones provided in Table 1, so no significant changes from the statistical point of view are obtained.  The test speaks out in favor of the robustness of the reported results.


\section{Relative size of the vacuum fluctuations}

Relation \eqref{eq:ExtraRelation} allows us to estimate the size of the vacuum energy density perturbations at deep subhorizon scales. According to the background formulas \eqref{eq:contBackground} and \eqref{eq:rhoM}, for the RVM we easily find
\begin{equation}\label{eq:psinuH}
\psi=-\frac{\dot{\rho}_m+3\mathcal{H}\rho_m}{\rho_m}=-\left(3+a\frac{\rho_m'}{\rho_m}\right)\mathcal{H}=3\nu\mathcal{H}\,,
\end{equation}
where a prime indicates differentiation with respect to the scale factor and, as indicated above, we assume that for the discussion on density perturbations we can entirely neglect the radiation component.
Using the above formula and the perturbed continuity equation \eqref{eq:ContinuityNewton} in \eqref{eq:ExtraRelation} one can check that $\frac{\delta\rho_\Lambda}{\delta\rho_m}\propto\nu\left(\frac{\mathcal{H}}{k}\right)^2$, see Figs 1-2 and the remaining discussion in this section. Therefore, the vacuum energy density perturbations are very much suppressed at scales deep inside the horizon ($k\gg \mathcal{H}$). Actually, since the fitting results in Table 1 tell us that cosmological observations prefer values of $\nu$ of order $\mathcal{O}(10^{-3})$,  this  helps to suppress even more the vacuum density fluctuation $\delta\rho_\Lambda$ with respect to the material one. As expected, we recover the $\Lambda$CDM result, i.e. $\delta\rho_\Lambda=0$, when we set $\nu=0$.
\begin{figure}
\includegraphics[scale=0.59]{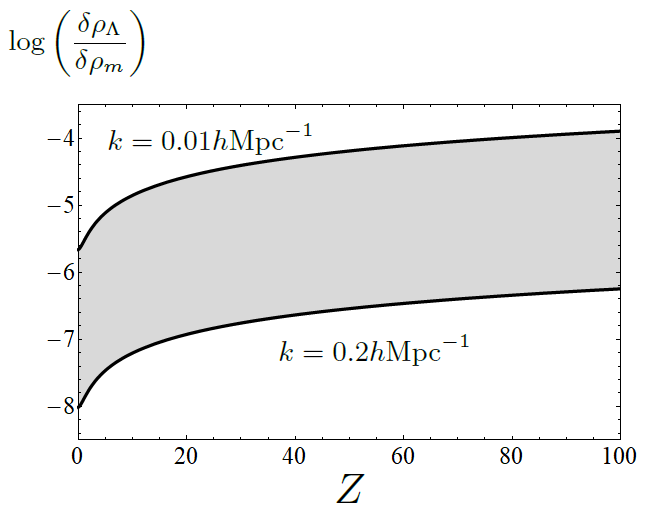}
\caption{Logarithm of the ratio of density perturbations of vacuum and matter  as a function of the redshift and for the same comoving wave numbers of Fig. 1. They are again inside the gray band.  The effect of vacuum energy perturbations is enhanced at large scales, but even for the largest comoving scales of interest it is negligible in front of the corrections considered in Eq. \eqref{eq:DensityContrastEq}, see the discussion in Sect. 4.
\label{fig:PertComparison}}
\end{figure}
Despite the $\delta\rho_\CC$ suppression, it is noticeable that the dynamical nature of  $\rL$ enables the formation of some structure in the vacuum sector at subhorizon scales, something that is strictly denied in the $\Lambda$CDM model. In the RVM the small clustering of the vacuum can be enhanced for larger and larger values of the vacuum parameter $\nu$, if they would be allowed. This is reasonable, since a defect or an excess in the matter distribution must generate also a departure from uniformity of the vacuum energy density, just because both components are directly interacting with each other. In the $\Lambda$CDM one deals with a strictly constant vacuum energy density, but in principle one could think that although there is no direct exchange of energy between matter and vacuum at the background level, the vacuum perturbations could affect the matter ones through the gravitational potential in Poisson's equation. It turns out not to be the case, since vacuum perturbations in the $\Lambda$CDM are strictly zero. This can be automatically inferred from \eqref{eq:split2} after setting $\mathcal{B}=0$.

Let us remark that we have initially begun with three independent equations, \eqref{eq:PoissonNewton}-\eqref{eq:EulerNewton}, and 4 unknown perturbed functions, $\delta\rho_m$, $\delta\rho_\Lambda$, $\Phi$ and $v_m$. By providing solid physical arguments we have motivated an additional relation between $\delta\rho_\Lambda$ and $v_m$, \eqref{eq:ExtraRelation}, which has allowed us to retrieve the standard Euler equation \eqref{eq:EulerNewton3}. This will also let us to close the system of perturbed equations in a consistent way and finally find the equation that governs the evolution of the matter density perturbations. But before doing this, it is illuminating to provide some details on computing the above mentioned ratio $\delta\rho_\Lambda/\delta\rho_m$ at subhorizon scales to within linear approximation in the small vacuum parameter $\nu$.  Such is, of course, also the parameter controlling the strength of the vacuum-matter interaction. The calculation of $\delta\rho_\Lambda/\delta\rho_m$ can be easily done by using Eq. \eqref{eq:ContinuityNewton} and the relation \eqref{eq:ExtraRelation}:
\begin{equation}\label{eq:Ratio1}
\frac{\delta\rho_\Lambda}{\delta\rho_m}=\frac{\psi}{k^2}\frac{\dot{\delta}_m}{\delta_m}+\mathcal{O}(\nu^2)\,,
\end{equation}
where $\delta_m=\delta\rho_m/\rho_m$ is the so-called matter density contrast and $\mathcal{O}(\nu^2)$ encapsulates all the explicit terms of second or higher order in $\nu$, i.e. those higher order corrections that are not implicit in the first term of the {\it r.h.s.} of \eqref{eq:Ratio1}. Equivalently, the last relation can also be rewritten as
\begin{equation}\label{eq:Ratio2}
\frac{\delta\rho_\Lambda}{\delta\rho_m}=-a\left(\frac{\mathcal{H}(a)}{k}\right)^2\frac{f(a)}{\rho_m(a)}\frac{d\rho_\Lambda}{da}+\mathcal{O}(\nu^2)\,,
\end{equation}
with
\begin{figure*}
\includegraphics[scale=0.7]{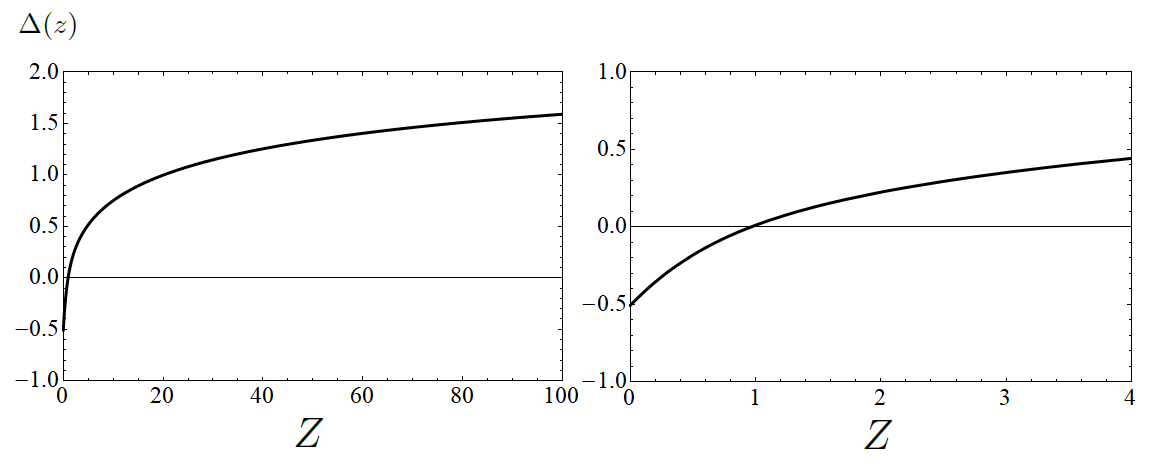}
\caption{{\it Left plot:} Evolution of the percentage difference of the matter density contrast in the RVM with respect to  the $\Lambda$CDM, $\Delta (z)$, as defined in \eqref{eq:RelDiff}, in the redshift range $0\leq z\leq 100$; {\it Right plot:} The same, but in the smaller redshift range $0\leq z\leq 4$, just to show the region where $\Delta(z)$ becomes negative, near the present time.  It is crystal-clear that such differences are always smaller than  $2\%$ in absolute value.
\label{fig:RelDiff}}
\end{figure*}
%
\begin{equation}\label{eq:growthrate}
f(a)=\frac{d\ln\delta_m}{d\ln\,a}
\end{equation}
the growth rate. For the RVM the ratio (\ref{eq:Ratio2}) can be cast in a compact form. Taking into account that the matter density   in the MD epoch for this model is given by the first term on the \textit{r.h.s.} of (\ref{eq:rhoM}) and that the corresponding  vacuum energy density in the same epoch reads
\begin{equation}\label{eq:rLMDE}
  \rho_\CC(a) = \rLo + \frac{\nu\,\rho_{m0}}{1-\nu}\left(a^{-3(1-\nu)}-1\right)\,,
\end{equation}
we find:
\begin{equation}
\frac{\delta\rho_\Lambda}{\delta\rho_m}=3\nu\left(\frac{\mathcal{H}(a)}{k}\right)^2f(a) +\mathcal{O}(\nu^2)\,.
\end{equation}
We note that $f(a)$ is a monotonic function that is around $0.5$ at present and saturates to $f\simeq 1$ at $z\simeq 1$ for models with a well-defined $\Lambda$CDM limit (cf. G\'omez-Valent  \& Sol\`a  2015).  This proves our contention that $\frac{\delta\rho_\Lambda}{\delta\rho_m}\propto\nu\left(\frac{\mathcal{H}}{k}\right)^2$.  Taking into account that $\nu=\mathcal{O}(10^{-3})$ , we find
\begin{equation}
\frac{\delta\rho_\Lambda}{\delta\rho_m}\bigg\rvert_{z=0}\sim\left(\frac{H_0}{k}\right)^2\cdot\mathcal{O}(10^{-3})\,,\end{equation}
\begin{equation}
\frac{\delta\rho_\Lambda}{\delta\rho_m}\bigg\rvert_{z=100}\sim\left(\frac{H_0}{k}\right)^2\cdot\mathcal{O}(10^{-1})\label{eq:N2}\,,
\end{equation}
where $H_0\simeq 3.336\times 10^{-4}h{\rm Mpc}^{-1}$, and where we have used $\mathcal{H}^2=H^2a^2=H^2/(1+z)^2\simeq H_0^2\Omega_m(1+z)$ at high redshift within the MD epoch.  Recalling that the observational data concerning the linear power spectrum lie in the approximate comoving wave number range $0.01\,h{\rm Mpc}^{-1}\lesssim k\lesssim 0.2\,h{\rm Mpc}^{-1}$, or equivalently, $0.002\lesssim H_0/k\lesssim 0.03$, we find that the above ratios of the energy density perturbations sit in the following intervals:
\begin{equation}\label{eq:bounds1}
10^{-9}\lesssim \frac{\delta\rho_\Lambda}{\delta\rho_m}\bigg\rvert_{z=0}\lesssim 10^{-6}\,,
\end{equation}
\begin{equation}\label{eq:bounds2}
10^{-7}\lesssim\frac{\delta\rho_\Lambda}{\delta\rho_m}\bigg\rvert_{z=100}\lesssim 10^{-4}\,,
\end{equation}
\noindent where the lower bounds in these expressions have been obtained using the lowest value of $H_0/k$ we have observational access to, i.e. $H_0/k\sim 0.002$, and the upper bound with the largest accessible value of $H_0/k\sim 0.03$. In Fig. 1 we show that these modes are deeply inside the horizon in the epoch of interest, viz. from $z\sim 100$ up to the present time. Therefore, we are fully legitimated to apply the subhorizon approximation for the relevant modes. The bounds \eqref{eq:bounds1}-\eqref{eq:bounds2} completely agree with the results presented in Fig. 4 of (G\'omez-Valent, Karimkhani \& Sol\`a 2015), in which we studied the effect of DE perturbations at subhorizon scales in the context of the so-called $\mathcal{D}$-class of dynamical DE models. By direct inspection of \eqref{eq:bounds1}-\eqref{eq:bounds2} and Fig. 2 it becomes evident that $\delta\rho_m\gg\delta\rho_\Lambda$ at subhorizon scales. In principle, this allows us to neglect the vacuum energy density perturbations and its derivatives in front of the matter ones. But before accepting this we must still address one more question. Are the effects that come from not neglecting $\delta\rho_\Lambda$ in front of $\delta\rho_m$ of the same order of magnitude as those coming from the fact of having the pure background extra effect $\psi\ne 0$, see Eq.\,(\ref{eq:psinuH}), associated to the vacuum time evolution? If this were the case, then it would be a  fairer approximation to our correction to the standard perturbations equations in the presence of vacuum dynamics to keep $\delta\rho_\Lambda\ne 0$. We will see now that it is not the case.
We have checked before that $\delta\rho_m\gg\delta\rho_\Lambda$.
Let us therefore neglect the vacuum perturbations in front of the matter ones now. Then, equations \eqref{eq:PoissonNewton} and \eqref{eq:ContinuityNewton} can be rewritten in a simpler way, as follows:
\begin{align}
k^2\Phi &= -4\pi G a^2\rho_m\delta_m\,,\label{eq:PoissonNewton2}\\
\dot{\delta}_m+\psi\delta_m&=k^2v_m\,,\label{eq:ContinuityNewton2}
\end{align}
The continuity equation \eqref{eq:ContinuityNewton2} is modified with an extra term, i.e. $\psi\delta_m$, with respect to the $\Lambda$CDM case because now there is an injection/extraction of energy in the matter sector caused by the decay/increase of the vacuum energy density.

Combining \eqref{eq:EulerNewton3}, \eqref{eq:PoissonNewton2} and \eqref{eq:ContinuityNewton2} the following second order equation for the matter density contrast is obtained:
\begin{equation}\label{eq:DensityContrastEq}
\ddot{\delta}_m+\dot{\delta}_m(\mathcal{H}+\psi)+\delta_m(-4\pi G a^2\rho_m+\dot{\psi}+\psi\mathcal{H})=0\,.
\end{equation}
As expected, in the absence of matter-vacuum interaction, i.e. $\psi=0$, we retrieve the equation of the standard $\Lambda$CDM model. To solve Eq. \eqref{eq:DensityContrastEq} we need to set two initial conditions, for simplicity at a time at which the expansion is fully matter-dominated. It could be,  say at  $a_i\sim 1/100$, but it is even better to chose $a_i$ around $1/5$ or $1/10$ in order to maximally suppress the theoretical error associated to the use of the subhorizon approximation (cf. Fig. 1), which at these values of the scale factor is lower than $0.4\%$ for all the modes $k$ of interest. At these redshifts the universe is still strongly matter-dominated:  $\Omega_m a^{-3}\gg\Omega_\CC$.

In the MD era the density contrast can be computed analytically. Changing from conformal time to the scale factor through the chain rule $d/d\eta=a\mathcal{H}d/da=a^2Hd/da$, and then using the simplified form for $H$ in the matter epoch (given by the first two terms on the \textit{r.h.s} of (\ref{eq:H2RVM})),  the perturbations equation (\ref{eq:DensityContrastEq}) boils down to
\begin{equation}\label{diffeqDaRVM}
\delta''_m + \frac{3}{2a}(1+3\nu)\delta'_m - \left(\frac{3}{2}-3\nu-\frac{9}{2}\,\nu^2\right)\frac{\delta_m}{a^2} = 0\,.
\end{equation}
The (exact) growing mode solution of this equation is the power-law   $\delta_m(a)=a^{1-3\nu}$, so we can use this relation to set the initial conditions for $\delta_m$ and its first derivative.

Eq. \eqref{eq:DensityContrastEq} was obtained by assuming a perfectly homogeneous dynamical vacuum energy density at deep subhorizon scales. The corrections introduced by the non-standard terms can easily be evaluated in the RVM, even analytically.  Their relative weights with respect to the standard parts in (\ref{eq:DensityContrastEq}) read as follows. On the one hand we have already indicated in (\ref{eq:psinuH}) that
\begin{equation}\label{eq:psioverH}\frac{\psi}{\mathcal{H}}=3 \nu\,,\end{equation}
and similarly we obtain for the other  non-standard  terms:
\begin{equation}
\frac{\psi\mathcal{H}}{4\pi G a^2\rho_m}=2\nu\left(1+\frac{\rho_\Lambda}{\rho_m}\right)\lesssim\frac{20}{3}\nu\,,\end{equation}
\begin{equation}\label{eq:dotpsi}
\left|\frac{\dot{\psi}}{4\pi G a^2\rho_m}\right|=\nu\left|2\frac{\rho_\Lambda}{\rho_m}-1\right|\lesssim \frac{11}{3}\nu\,,
\end{equation}
where we recall that $\nu>0$ from our fit.  The upper bounds in the above expressions correspond to the ratios near our time, for which $\rho_\CC/\rho_m\simeq \OL/\Om\simeq 7/3$. In the past, e.g. deep in the MD epoch,  the bound is of course tighter since at that time $\rho_\CC/\rho_m\ll 1$.  To reach e.g. the bound (\ref{eq:dotpsi}) we can use $\psi=3\nu\mathcal{H}=3\nu a H$ from  (\ref{eq:psioverH}) and the leading terms of (\ref{eq:H2RVM}) in the MD epoch. We obtain
\begin{eqnarray}\label{eq:dotpsi2}
  \dot{\psi}&=&3\nu H_0^2 a^2 \left(E^2+\frac{a}{2}\frac{dE^2}{da}\right)=3\nu H_0^2 a^2 \left(E^2-\frac32\frac{\rho_m}{\rco}\right)\nonumber\\
  &=&\frac32\,\frac{\nu H_0^2 a^2}{\rco}\left(2\rL-\rho_m\right)\,,
\end{eqnarray}
where use has been made of  $E^2=\left(\rho_m+\rho_{\CC}\right)/\rco$.
The relative corrections computed above are seen to be of order $\mathcal{O}(10^{-3}-10^{-2})$ in all cases, and are therefore much larger than the corrections associated to the clustering of the vacuum energy, which proves to be lower than $\mathcal{O}(10^{-4})$ for the modes under study and for redshifts lower than $z\sim 100$, see equations (\ref{eq:bounds1}) and (\ref{eq:bounds2}).

We may crosscheck the above result by comparing the value of the matter density contrast, which evolves from $\delta_m(z\sim 100)=\mathcal{O}(10^{-2})$ to $\delta_m(z=0)=\mathcal{O}(1)$, with the values of $\delta\rho_\Lambda/\rho_m$, which are of order $\mathcal{O}(10^{-9}-10^{-6})$ for any accessible scale at $z\lesssim 100$. The latter follow from $\delta\rho_\Lambda/\rho_m=\delta_m\delta\rho_\Lambda/\delta\rho_m$ and the bounds on the ratio $\delta\rho_\Lambda/\delta\rho_m$ \eqref{eq:bounds1}-\eqref{eq:bounds2}. Thus, from the former analysis we conclude that we can safely neglect the vacuum energy density perturbations and use Eq.\,\eqref{eq:DensityContrastEq} to study the evolution of the matter density contrast at deep subhorizon scales.

Let us now elucidate what are the relative differences between the matter density contrast $\delta_m(z)$ as obtained from Eq. \eqref{eq:DensityContrastEq} (with $\psi\ne 0$) and the standard one in the  $\Lambda$CDM model (corresponding to   $\psi=0$ ).  In the last case we will denote the resulting density contrast as $\tilde{\delta}_m(z)$. The percentage differences
\begin{equation}\label{eq:RelDiff}
\Delta(z)\equiv 100\cdot\frac{\delta_m(z)-\tilde{\delta}_m(z)}{\tilde{\delta}_m(z)}
\end{equation}
are shown in Fig.\,3. We can read-off from it that the corrections introduced in Eq. \eqref{eq:DensityContrastEq} by the terms that are proportional to $\psi$ or its time derivative are lower than $2\%$ in absolute value. Despite being small, it is important to take them into account, since the fitting results are already sensitive to them. Note  that some RSD data points, e.g. those from Gil-Mar\'in et al. 2017, have a relative error of only few percent, and in the near future the error  will decrease  in some cases below  $1\%$ (Weinberg et al. 2013).

Owing to the importance of the subject, let us further dwell upon the perturbations equations in the presence of vacuum dynamics. It turns out that the same equation \eqref{eq:DensityContrastEq} can also be derived using a source 4-vector $Q_\mu=Q u_\mu$, with $Q=-\mathring{\rho}_\Lambda$ (the circle denotes hereafter a derivative with respect to the cosmic time) and $u_\mu=\bar{u}_\mu+\delta u_\mu$ the perturbed 4-velocity of the matter fluid in natural units ($c=1$), where $\bar{u}_\mu=(a,\vec{0})$ and $\delta u_\mu=a\left(\Phi,-\vec{v}_m\right)$. The use of  $Q_\mu$ ensures the automatic fulfillment of \eqref{eq:ExtraRelation} and the usual Euler equation \eqref{eq:EulerNewton3}. Let us see this more in detail. Due to the Bianchi identity, we find
\begin{equation}
\nabla^\mu (T^{\rm m}_{\mu\nu}+T^{\rm \Lambda}_{\mu\nu})=0\,,
\end{equation}
$T^{\rm m}_{\mu\nu}$ and $T^{\rm \Lambda}_{\mu\nu}$  being the matter and vacuum energy-momentum tensors, respectively. We can split this equation in two parts by means of $Q_\mu$,
\begin{equation}\label{eq:splitQ}
\nabla^\mu T^{\rm m}_{\mu\nu}\equiv Q_\nu\qquad\nabla^\mu T^{\rm \Lambda}_{\mu\nu}\equiv -Q_\nu\,.
\end{equation}
The perturbed source vector yields,
\begin{equation}\label{eq:deltaQ}
\delta Q_\mu=\delta Q\bar{u}_\mu+Q\delta u_\mu=(a\delta Q+a\Phi Q,-aQ\vec{v}_m)\,.
\end{equation}
Let us now perturb the first equation of \eqref{eq:splitQ} and substitute (\ref{eq:deltaQ}) on its \textit{r.h.s.}:
\begin{equation}\label{eq:deltaTmn}
\delta\left(\nabla^{\mu} T_{\mu\nu}^m\right)=(a\delta Q+a\Phi Q,-aQ\vec{v}_m)\,,
\end{equation}
Writing out the spatial component ($\nu=i$)  along the lines of  (\ref{eq:Euler1}) and defining $\delta Q_i\equiv \partial_i\delta V$, with $\delta V=-aQv_m=a\mathring{\rho}_\CC v_m=\dot{\rho}_\Lambda v_m$, we find
\begin{equation}
-\rho_m\left(\dot{v}_m+\mathcal{H}v_m+\Phi+\psi v_m\right)={\delta V}\,,
\end{equation}
and hence
\begin{equation}
\dot{v}_m+\mathcal{H}v_m+\Phi=-\frac{\delta V}{\rho_m}-\psi v_m=\frac{-\delta V+\dot{\rho}_\Lambda v_m}{\rho_m}=0\,,
\end{equation}
so the usual Euler equation is retrieved.  This shows that this alternative procedure is equivalent to use the setting (\ref{eq:ExtraRelation}) on Eq.\,(\ref{eq:EulerNewton2}). We can proceed similarly with the time component, i.e. the perturbed continuity equation, and we find
\begin{equation}
\dot{\delta}_m-k^2v_m+\psi \delta_m=\frac{\delta Q_0}{\rho_m}=\frac{a}{\rho_m}(\delta Q+\Phi Q)\,.
\end{equation}
The fact that we have already neglected the term proportional to $\dot{\Phi}$, originally present in \eqref{eq:ContinuityOriginal}, is completely justified at subhorizon scales (cf. Eq. \eqref{eq:PoissonNewton2}).  The last equation can be rewritten as follows:
\begin{equation}
\dot{\delta}_m-k^2v_m+\psi \delta_m=-\left(\frac{\delta\dot{\rho}_\Lambda+\Phi \dot{\rho}_\Lambda}{\rho_m}\right)\,.
\end{equation}
As we have checked before, vacuum fluctuations are negligible at subhorizon scales as compared to matter fluctuations, so the first term on the {\it r.h.s.} can be clearly neglected when compared with the first term of the {\it l.h.s.} The second term on the {\it r.h.s.} can be neglected  too, but for a different reason. Because of the Poisson equation \eqref{eq:PoissonNewton2}, $\Phi\propto G a^2 \delta\rho_m/k^2$, at subhorizon scales. It follows that
\begin{equation}\label{eq:ratiosubhorizon}
  \frac{\Phi \dot{\rho}_\Lambda/\rho_m}{\psi \delta_m}\propto \frac{G a^2\rho_m}{k^2}\propto\frac{\mathcal{H}^2}{k^2}\ll1
\end{equation}
and therefore we are allowed to neglect the second term of the {\it r.h.s.} as well, so we meet once more the continuity equation \eqref{eq:ContinuityNewton2}.


\section{Density perturbations with vacuum dynamics at subhorizon scales in the synchronous  gauge}\label{sect:SynchronousGauge}
To hammer this important point home from a different perspective, let us now obtain the same equation \eqref{eq:DensityContrastEq} for the matter density contrast at low scales using the synchronous gauge (Ma \& Bertschinger 1995). This will help to further illustrate the robustness of our previous results. In the synchronous gauge the perturbed FLRW metric reads:
\begin{equation}
ds^2=dt^2-(a^2\delta_{ij}-h_{ij})dx^i dx^j\,.
\end{equation}
In this case the three basic perturbations equations read (see e.g. Grande, Pelinson \& Sol\`a 2009):
\begin{equation}\label{eq:syn1}
\mathring{\hat{h}}+2H\hat{h}=8\pi G\sum_{i=\Lambda,m}(\delta\rho_i+3\delta p_i)\,,
\end{equation}
\begin{equation}\label{eq:syn2}
\sum_{i=\Lambda,m}\delta\rho_i+(\rho_i+p_i)\left(\theta_i-\frac{\hat{h}}{2}\right)+3H(\delta\rho_i+\delta p_i)=0\,,
\end{equation}
\begin{equation}\label{eq:syn3}
\sum_{i=\Lambda,m} \mathring{\theta}_i(\rho_i+p_i)+\theta_i\left[\mathring{\rho}_i+\mathring{p}_i+5H(\rho_i+p_i)\right]=\frac{k^2}{a^2}\sum_{i=\Lambda,m}\delta p_i\,,
\end{equation}
in which $\hat{h}=\frac{\partial}{\partial t}\left(\frac{h_{ii}}{a^2}\right)=-\mathring{h}$  ($h_{ii}$ being the trace) is one of the two scalar modes of the spatial metric perturbations in this gauge, and $\theta_i$ is the covariant derivative of the i$_{th}$ component velocity perturbation, i.e. $\theta=\nabla_\mu\delta u^\mu$, with $\delta u^\mu=\frac{1}{a}\left(0,\vec{v}\right)$ and $\vec{v}=\vec{\nabla}v$. We assume that the vacuum has no peculiar velocity and, again, that $\delta\rho_m\gg\delta\rho_\Lambda$ and $\mathring{\delta\rho_m}\gg\mathring{\delta\rho_\Lambda}$. In Fourier space we obtain:

\begin{figure*}
\includegraphics[scale=0.6]{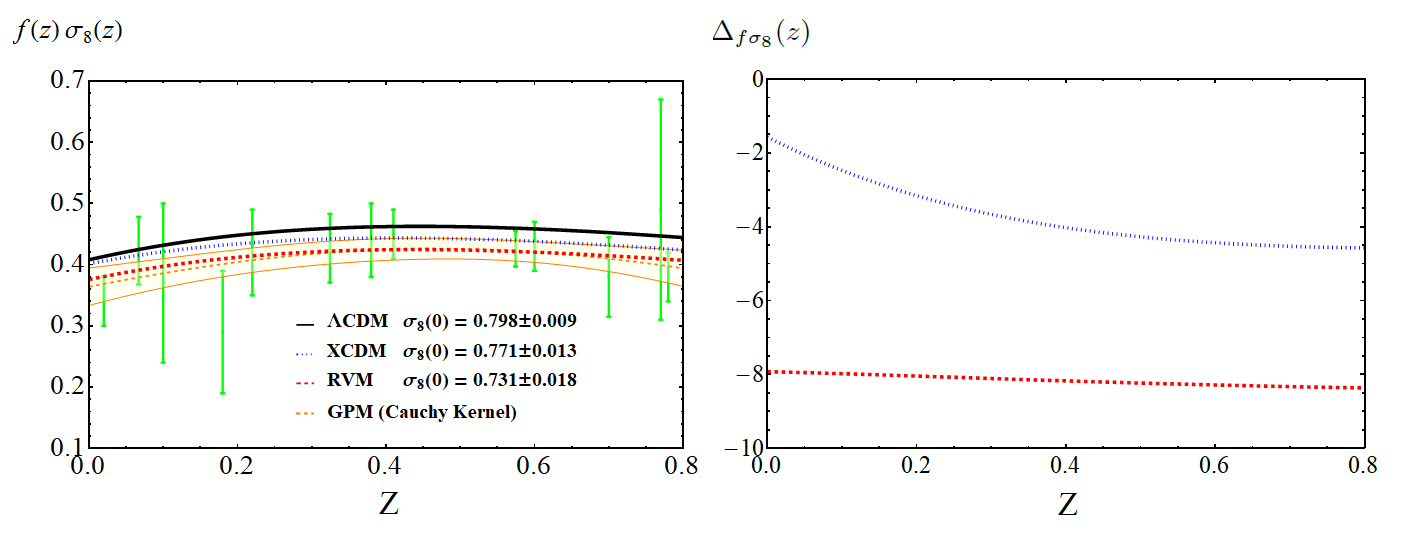}
\caption{{\it Left plot:} The weighted growth rate for the $\Lambda$CDM, the XCDM and the RVM, obtained by using the best-fit values of Table 1. The values of $\sigma_8$ that we obtain for these models are also indicated. We also plot the reconstructed $f(z)\sigma_8(z)$ curve and its $1\sigma$ uncertainty band, both obtained by using the observational data (depicted in green) and the Gaussian processes method (GPM) with Cauchy's kernel, see e.g. (Seikel, Clarkson \& Smith 2012) and references therein. Almost identical results are obtained using alternative kernels as the Gaussian or the Mat\'ern ones. This is to show the preference of the data for lower values of this LSS observable; {\it Right plot:} The relative (percentage) difference of the weighted growth rate with respect to the concordance model, $\Delta_{f\sigma_8}$, as defined in \eqref{eq:Deltafsigma8}.
\label{fig:fsigma8}}
\end{figure*}

\begin{align}
\mathring{\hat{h}}+2H\hat{h}&=8\pi G \delta\rho_m\,,\label{eq:syn1b}\\
      \mathring{\delta\rho_m}+\rho_m\left(\theta_m-\frac{\hat{h}}{2}\right)+3H\delta\rho_m&=0\,,\label{eq:syn2b}\\
       \rho_m\mathring{\theta}_m+(\mathring{\rho}_m+5H\rho_m)\theta_m&=-\frac{k^2}{a^2}\delta\rho_\Lambda\label{eq:syn3b}\,.
\end{align}
The last equation can be cast as follows,
\begin{equation}\label{eq:thetaEq}
\mathring{\theta}_m+\theta_m\left(\bar{\psi}+2H\right)=-\frac{k^2}{a^2}\frac{\delta\rho_\Lambda}{\rho_m}\,.
\end{equation}
with $\bar{\psi}=-\mathring{\rho}_\Lambda/\rho_m$, which is the analog  in cosmic time of   ${\psi}=-\dot{\rho}_\Lambda/\rho_m$ defined in the previous sections with conformal time.
In coordinate space,
\begin{equation}
\theta_m=\nabla_\mu\delta u^\mu=\nabla_\mu(g^{\mu\nu}\delta u_\nu)=g^{ij}\partial_j \delta u_i+\mathcal{O}(2)\,,
\end{equation}
or
\begin{equation}
\theta_m=\frac{-\vec{\nabla}\cdot(-a\vec{v}_m)}{a^2}+\mathcal{O}(2)=\frac{\nabla^2v_m}{a}+\mathcal{O}(2)\,,
\end{equation}
where $\mathcal{O}(2)$ refers to second order perturbations. In momentum space,
\begin{equation}
\theta_m=-\frac{k^2}{a}v_m+\mathcal{O}(2)\,.
\end{equation}
Substituting this in \eqref{eq:thetaEq} the $k$-dependence cancels and we find:
\begin{equation}\label{eq:mathringv}
\mathring{v}_m+Hv_m+v_m\bar{\psi}=\frac{1}{a}\frac{\delta\rho_\Lambda}{\rho_m}\,.
\end{equation}
Equation (\ref{eq:mathringv}) is the momentum conservation equation for the matter particles in the synchronous gauge. As in the Newtonian gauge, we do not want this equation to be modified with respect to the $\Lambda$CDM one, where $\mathring{v}_m+Hv_m=0$. Thus, we impose $\delta\rho_\Lambda=\bar{\psi }v_m\rho_m$, which is formally the same kind of relation that we have obtained in the Newtonian gauge. But we must still fix the residual gauge freedom characteristic of the synchronous gauge  (Ma \& Bertschinger 1995).
 In the Newtonian gauge  we have the Bardeen potentials, $\Phi$ and $\Psi$, with $\Psi=-\Phi$ for perfect fluids. In the synchronous gauge, one can see that $\Phi$ is  absorbed in the trace of $h_{ij}$ (see e.g. Grande, Pelinson \& Sol\`a 2009). The second scalar mode existing in this gauge (which is contained in the longitudinal part of the metric) still gives us a residual gauge freedom that we are entitled to use. With it we can make a suitable choice that encompasses the situation in the $\Lambda$CDM, in which  $\bar{\psi}=0$, and where the residual gauge freedom can be fixed by setting $v_m=0$. Similarly, in the DVMs in general, and in the RVM in particular, we can exploit the remaining gauge freedom left by imposing that the peculiar velocity of matter particles is zero, i.e. once more the comoving frame  condition $v_m=0$  (see e.g. Wang, Wands, Xu, De-Santiago \& Hojjati 2013; Wang, Wands, Zhao \& Xu 2014). This setting automatically leads to $\delta\rho_\Lambda=0$  from (\ref{eq:mathringv}). It follows that the vacuum energy perturbations vanish in the comoving frame of matter, and as a result the dependence on $k$ can be seen to drop from all the above equations.

In the the Newtonian gauge we have a qualitatively different picture, which can nevertheless be made quantitatively coincident under appropriate conditions. The presence of the potential $\Phi$ (which is nothing but the Newton potential in the nonrelativistic limit) is inherently associated to $k$ through the Poisson equation.  Such $k$-dependence, however, disappears from the resulting matter density perturbations equation \eqref{eq:DensityContrastEq}, but only if we work at scales deeply below the horizon. In this sense the conformal Newtonian gauge is more physical since it tracks continuously the $k$-dependence and informs us on the physical conditions under which such dependence becomes negligible.

In the synchronous gauge there is no Newtonian limit; notwithstanding,  the same physical result for the material density perturbations ensues if we use the mentioned comoving setting for the peculiar velocities of the matter particles, viz.
$v_m=0$, which implies $\delta\rho_{\CC}=0=\theta_m$.  Indeed,  upon appropriate manipulation of the remaining equations \eqref{eq:syn1b} and \eqref{eq:syn2b} we can derive the following second order differential equation for the matter perturbations within the synchronous gauge and in cosmic time:
\begin{equation}\label{diffeqD}
\ringring{\delta}_m+\left(2H+\bar{\psi}\right)\,\mathring{\delta}_m-\left(4\pi
G\rmr-2H{\bar{\psi}}-\mathring{\bar{\psi}}\right)\,\delta_m=0\,.
\end{equation}
This is the final equation for matter perturbations well below the horizon.  At this point we can jump once more back to conformal time with the help of simple relations such as  $\mathring{\delta}= \dot{\delta}/a$, $\ringring{\delta}=(\ddot{\delta}-\mathcal{H}\dot{\delta})/a^2$, as well as $\bar{\psi}=\psi/a$ and $\mathring{\bar{\psi}}=(\dot{\psi}-\mathcal{H}\psi)/a^2$. In this way  we arrive once more to Eq.\,\eqref{eq:DensityContrastEq}. The final matter perturbations equation in both gauges is therefore the same,  provided we consider  scales sufficiently small as compared to the horizon. This result, which was well known for the $\CC$CDM (Ma \& Bertschinger 1995) has been proven here to hold good for the DVMs too.

Let us remark that even if we would not fix the residual gauge condition by picking  a comoving frame for the matter particles, namely  if we instead use $\delta\rho_\Lambda=\bar{\psi} v_m\rho_m\neq 0$ and solve $\mathring{v}_m+Hv_m=0$, we find that the velocity potential reads $v_m(a)=v_m(a_i)a_i/a$, where $a_i$ is some initial value of the scale factor far in the past but still in the MD epoch. This solution is a decaying mode and therefore it is expected to have a completely negligible effect on the matter energy density perturbations. In practice, we can set $v_m(a)\simeq 0$, what directly leads to $\delta\rho_\Lambda\simeq 0$ because the latter is not only proportional to $v_m$, but also to $\bar{\psi}$,  so in practice the quantity $\delta\rho_\Lambda=-\mathring{\rho}_\CC v_m$  is additionally suppressed by the tiny time variation of $\rL$.

The fact that at scales deeply inside the horizon the differential equation that controls the evolution of the matter density perturbations is the same in both gauges is somehow expected since it is already so for the $\Lambda$CDM case  (see e.g. Ma \& Bertschinger 1995). In the Newtonian gauge one measures a non-zero matter longitudinal velocity field, whereas in the synchronous gauge the observer free-falls with the matter fluid flow. Therefore  in this gauge no peculiar velocity for matter particles is measured, although the observer can in principle detect differences between the baryon and dark matter peculiar velocities.  The concordance of the matter perturbations in the two gauges  is also a reflex that there is no significant scale-dependence (i.e. $k$-dependence) in the evolution of the density perturbations below the horizon, which means that all subhorizon modes evolve essentially alike, the corrections being of order $\mathcal{H}^2/k^2\ll1$.  Gauge differences, however, can be significant  for calculations involving very large scales, and of course  for super-horizon scales.  In that case, one must keep the appropriate $k$-dependence (as e.g. in the Newtonian gauge) or resort to a gauge-invariant formalism (Bardeen 1980; Kodama \& Sasaki 1984).  Let us also note that the reason why the physical discussion can be more transparent in the Newtonian gauge is because the time slicing in this gauge respects the isotropic expansion of the background. The synchronous gauge, instead, corresponds to free falling observers at all points (as previously indicated), what implies that its predictions are relevant only to length scales significantly smaller than the horizon, but at these scales it renders the same physics as the Newtonian gauge (Ma \& Bertschinger 1995; Mukhanov, Feldman \&  Brandenberger 1992). While all these facts have been known since long for the $\CC$CDM, here we have re-examined them in the context of dynamical vacuum models, and we have shown that the main features are preserved.

A short summary of the analysis presented in the last two sections is now in order. By imposing a fully consistent physical condition in both gauges or, equivalently, by choosing an appropriate interaction 4-vector $Q_\mu=Qu_\mu$ that ensures the setting in a covariant manner, we have shown the following two important results: (i) the vacuum energy density perturbations are definitely negligible at low (subhorizon) scales in front of the matter ones; and (ii) in both considered gauges (Newtonian and synchronous) we find the same modified law (i.e. different from the $\CC$CDM one)  which governs the matter density contrast in the presence of vacuum dynamics, viz. Eq. \eqref{eq:DensityContrastEq}, or equivalently Eq.\,(\ref{diffeqD}). In the absence of that dynamics, the modified equation reduces to the standard $\CC$CDM one.


\section{Weighted growth rate and matter power spectrum}

The weighted growth rate, $f(z)\sigma_8(z)$, has become one the most important LSS observables because of its ability to constrain cosmological models. One of its main advantages is that it is independent of the bias between the observed galaxy spectrum and the underlying (total) matter power spectrum (Guzzo et al. 2008; Song \& Percival 2009) and therefore it is protected from the side effects that might be introduced by the assumption of a particular fiducial cosmological model in the calculation of the bias factor $b(z)$. Nevertheless, these data points are not completely model-independent, since the observational teams must assume a specific fiducial model (the $\Lambda$CDM, for convenience) in order to infer cosmological distances from the measured redshifts. This model-dependence can be removed from the $f(z)\sigma_8(z)$ data points by e.g. rescaling them as in (Macaulay, Wehus \& Eriksen 2013; Nesseris, Pantazis \& Perivolaropoulos 2017). We have explicitly checked that, when applied, the mentioned correction has very low impact on the fitting results presented in Table 1. We find that all the numbers remain almost unaltered, as e.g. the values of $\chi^2_{\rm min}$ for the various models, which only undergo very mild corrections of $0.5\%$ at most; or the RVM parameter, which after the data rescaling reads $\nu=0.00162\pm0.00042$ and, therefore, keeps the very same level of significance ($\sim 3.85\sigma$) as the one shown in Table 1.

The weighted growth rate is given by the product of the growth rate $f(z)$, defined in \eqref{eq:growthrate}, and
\begin{equation}
\sigma_8(z)=\sigma_8\delta_m(z)/\delta_m(z=0)\,.
\end{equation}
In the left plot of Fig. 4 we show the theoretical curves of the weighted growth rate for the $\Lambda$CDM, the XCDM parametrization and the RVM. In the concordance model there is an obvious excess of power when compared with the other two scenarios, specially with the RVM. The exact relative difference with respect to the concordance model, defined as
\begin{equation}\label{eq:Deltafsigma8}
\Delta_{f\sigma_8}(z)\equiv100\cdot\frac{f(z)\sigma_8(z)\bigg\rvert_{{\rm Y}}-f(z)\sigma_8(z)\bigg\rvert_{\Lambda{\rm CDM}}}{f(z)\sigma_8(z)\bigg\rvert_{\Lambda{\rm CDM}}}\,,
\end{equation}
with Y = (XCDM, RVM), can be read off in the right plot of the same figure. It reaches a (negative) $2-4\%$ level in the XCDM and is enhanced up to $8-9\%$   (negative too) in the RVM.  Now, because the  $f(z)\sigma_8(z)$ data points lie some $\sim 8\%$  below the $\CC$CDM prediction (what is nothing more than the aforementioned $\sigma_8$-tension), the mentioned relative  differences allow the RVM to fit better the LSS data than the other two models under study. Such differences in $f(z)\sigma_8(z)$ actually agree with those that are found in the value of the $\sigma_8$ parameter (cf. the legend of the left plot in Fig. 4), which are around $3.4\%$ lower in the XCDM and $8.4\%$ lower in the RVM. This is probably telling us that the main source of the differences observed between the theoretical curves of $f(z)\sigma_8(z)$ come precisely from the predicted values of $\sigma_8$.

The main aim of the present and the next sections is to disentangle the origin of the above differences (\ref{eq:Deltafsigma8}) and to see which is the role played by the various parameters involved in the calculation of the weighted growth rate in the framework of both the XCDM and the RVM. It is particularly intriguing to understand how can the vacuum parameter $\nu$ in the RVM have such a great impact on the LSS predictions, taking into account that it is ``only'' of order $\mathcal{O}(10^{-3})$. The answer was advanced in (G\'omez-Valent \&  Sol\`a 2018) and here we will provide more details.

The following question should be addressed: Why are the induced changes with respect to the $\Lambda$CDM not of order $\nu$, as one could naively expect at linear order? It was shown in (G\'omez-Valent, Sol\`a \& Basilakos 2015) that in the non-linear perturbations regime the effect of a non-null $\nu$ can be very big, giving rise to differences in the prediction of the collapsed number of halos that can reach the $50\%$ level in some cases for typical values of $\nu\sim 10^{-3}$. This was studied in the context of an improved version of the Press-Schechter formalism (Press \& Schechter 1974), see (G\'omez-Valent, Sol\`a \& Basilakos 2015)  for details. The effects in the linear perturbations regime are not as big as the observed ones at the non-linear one, but are nevertheless  higher than first glance expectations. An explanation is therefore mandatory at this point, and we do offer it here in detail.
Before going on, let us write $\sigma_8(z)$ in a convenient way, which will allow us to better capture the physical information encoded in it:
\begin{equation}\label{eq:s88generalNN}
\sigma_8^2(z)=\delta_m^2(z)\int\frac{d^3k}{(2\pi)^3}\, P(k,\vec{p})\,\,W^2(kR_8)\,.
\end{equation}
Here $P(k,\vec{p})=P_0\,k^{n_s}T^2(k,\vec{p})$ is the linear matter power spectrum, $P_0$ is its normalization factor and $T(k,\vec{p})$ the matter transfer function, with $\vec{p}$ being the vector that contains the parameters of the model. Function  $P(k,\vec{p})$ gives the spectrum, i.e. the Fourier transform of the two-point correlation function of the primordial linear density field, whereas  $T(k,\vec{p})$  modulates the shape of the gravitational potential in the MD epoch for every mode. For the latter we have adopted the usual BBKS form (Bardeen, Bond, Kaiser \& Szalay 1986):
\begin{equation}\label{eq:BBKS}
\begin{array}{ll}
T(x) = &\frac{\ln (1+0.171 x)}{0.171\,x}\Big[1+0.284 x + (1.18 x)^2+\\
   & + \, (0.399 x)^3+(0.490x)^4\Big]^{-1/4}\,.
\end{array}
\end{equation}
Originally, $x=k/k_{eq}$,  where
\be\label{keqDef}
k_{eq}=a_{eq}H(a_{eq})
\ee
is the value of the comoving wavenumber at the equality scale $a_{eq}$ between matter and radiation densities: $\rho_r(a_{eq})=\rho_m(a_{eq})$. It is well-known that \eqref{eq:BBKS} does not incorporate the effects produced by the tightly coupled photo-baryon plasma before the decoupling time. The fight between pressure and gravity in this coupled system generates the baryon acoustic oscillations in the matter power spectrum at ``small'' scales, i.e. for $k>k_{eq}$. The baryon density effects can be introduced in \eqref{eq:BBKS} through the modified shape parameter $\tilde{\Gamma}$ (Peacock \& Dodds 1994; Sugiyama 1995) in $x=k/(k_{eq}\tilde{\Gamma})$, with
\be
\tilde{\Gamma}=e^{-\Omega_b-\sqrt{2h}\frac{\Omega_b}{\Omega_m}}\,.
\ee
Alternatively, one can use the transfer function provided in (Eisenstein \& Hu 1998) instead of the BBKS one. The former already includes the baryonic effects. We have checked that the use of the
alternative matter transfer function does not produce any significant change in our results, so we stick to the BBKS by incorporating the baryon effects through the shape parameter $\tilde{\Gamma}$, as explained above, since it is easier to deal with from an analytical point of view.

We remark that $k_{eq}$ is a model-dependent quantity, which departs from the $\CC$CDM expression in those models in which matter and/or radiation are governed by a nonstandard continuity equation in which matter exchanges energy with vacuum, such as e.g. in the RVM. For the concordance model and the XCDM parametrization, $k_{eq}$ has the simplest expression:
\begin{equation}\label{keqCCprev}
k^\CC_{eq} = H_0\,\Omega_m\sqrt{\frac{2}{\Omega_r}}\,.
\end{equation}
%
\begin{figure*}
\includegraphics[scale=0.45]{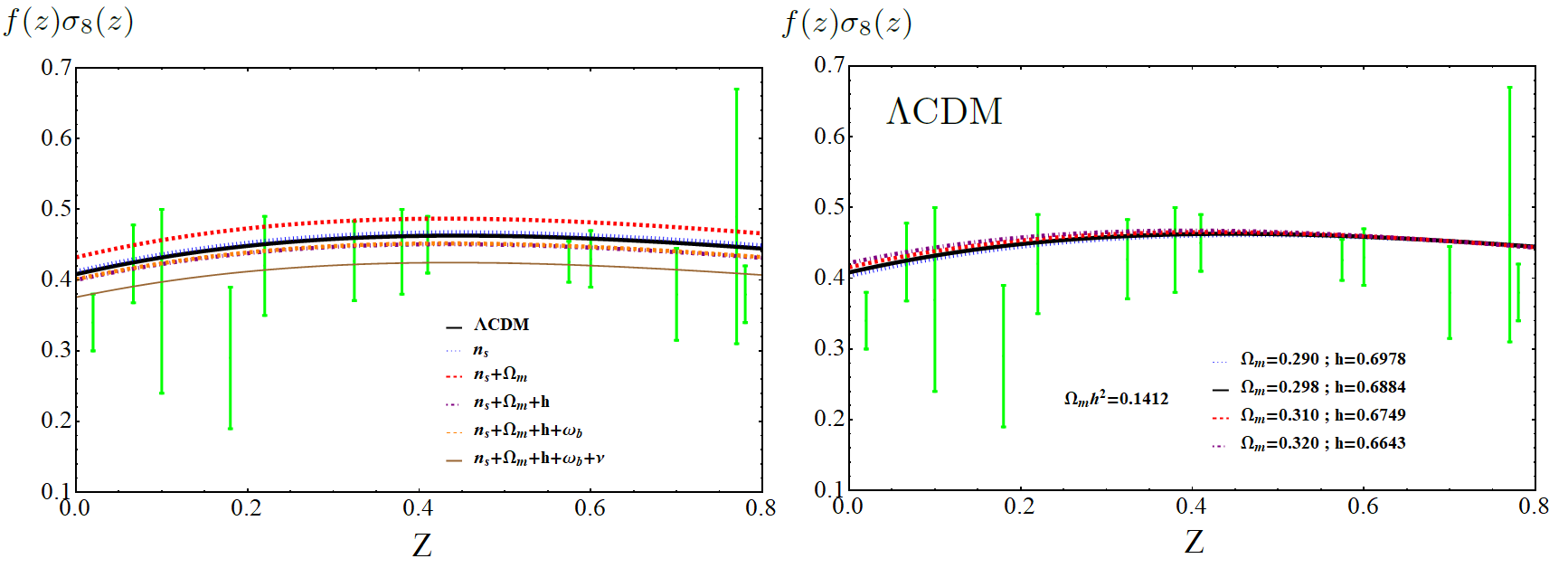}
\caption{{\it Left plot:} Reconstruction of the $f(z)\sigma_8(z)$  curve of the RVM from the $\Lambda$CDM one.  See the text in Sect. 7 for a detailed explanation; {\it Right plot:} Curves of $f(z)\sigma_8(z)$ obtained for the $\Lambda$CDM, with different values of $\Omega_m$ and $h$ satisfying the same relation $\omega_m\equiv\Omega_m h^2=0.1412$. This is to show the approximate degeneracy of the LSS results under modifications of $\Omega_m$ and $h$ that fully respect the strong constraint on $\omega_m$ coming from the CMB data, as explained in the text.
\label{fig:RVMcons}}
\end{figure*}
For the RVM, however, is not possible to find a formula as compact as \eqref{keqCCprev}. The corresponding expression for $a_{eq}$ is quite involved in this case. It follows from  $\rho_r(a_{eq})=\rho_m(a_{eq})$, in which $\rho_m(a)$ is the function (\ref{eq:rhoM}) and $\rho_r(a)$ is the standard density formula for conserved radiation.  We find:
\begin{eqnarray}\label{eq:aeqRVM}
\textrm{RVM}:\quad a_{eq}& =& \left[\frac{\Omega_r(1+7\nu)}{\Omega_m(1+3\nu)+4\nu\Omega_r}\right]^{\frac{1}{1+3\nu}}\\
   & =&\frac{\Omega_r}{\Omega_m}\left[1+4\nu-3\nu\ln\left(\frac{\Omega_r}{\Omega_m}\right)\right]+\mathcal{O}(\nu^2)\nonumber\,,
\end{eqnarray}
where in the last equality we have expanded up to linear terms in the small parameter $\nu$.
On the other hand, the Hubble rate at the equality time in the RVM can be extracted from (\ref{eq:H2RVM}). It reads, in very good approximation,
\begin{equation}\label{eq:E2eqRVM}
E^2(a_{eq})=\frac{\Omega_m^4}{\Omega_r^3}\left[2-30\nu+24\nu\ln\left(\frac{\Omega_r}{\Omega_m}\right)\right]+\mathcal{O}(\nu^2)\,.
\end{equation}
Thus, at linear order in the vacuum parameter, the wave number at equality, $k_{eq}$, takes the following form in the RVM:
\begin{equation}\label{eq:keqRVM}
k^{\rm RVM}_{eq}=k^\CC_{eq}\left[1-\frac{7\nu}{2}+3\nu\ln\left(\frac{\Omega_r}{\Omega_m}\right)\right]+\mathcal{O}(\nu^2)\,,
\end{equation}
where $k^\CC_{eq}$ is the standard value (\ref{keqCCprev}).
As expected, for $\nu=0$ we retrieve the values of $a_{eq}$ and $E^2_{eq}$ in the $\CC$CDM, i.e. $a_{eq}\to\Omega_r/\Omega_m$ and $E^2(a_{eq})\to 2\Omega_m^4/\Omega_r^3$, and also $k_{eq}\to k^\CC_{eq}$. Moreover, it is worth stressing at this point that although $\nu\sim \mathcal{O}(10^{-3})$, the relative change in $k_{eq}$ caused by it is not just of order $\nu$, but it is significantly enhanced owing to the large log up to roughly $3\nu|\ln ({\Omega_r}/{\Omega_m})|\sim 25\nu$, hence a result which is comfortably one order of magnitude larger than naively expected . This point will be important in the discussion of Sect. 7.

Function $W(kR_8)$ in  Eq.\,\eqref{eq:s88generalNN} is a top-hat smoothing function, which can be expressed in terms of the spherical Bessel function of order $1$, as follows:
\begin{equation}\label{eq:WBessel}
W(kR_8)=3\,\frac{j_1(kR_8)}{kR_8}=\frac{3}{k^2R_8^2}\left(\frac{\sin{\left(kR_8\right)}}{kR_8}-\cos{\left(kR_8\right)}\right)\,,
\end{equation}
with $R_8=8{h^{-1}}$ Mpc. In the fitting analysis of Sol\`a, G\'omez-Valent \& de Cruz P\'erez 2017d, from where we have taken the values of the various parameters (cf. Table 1), we have fixed the power spectrum normalization factor $P_0$ as follows,
\begin{equation}\label{eq:P0}
\begin{small}
P_0=\frac{\sigma_{8,\Lambda}^2}{\delta^2_{m,\Lambda}}\left[\int_{0}^{\infty} \frac{d^3k}{(2\pi)^3}k^{n_{s,\Lambda}}T^2(k,\vec{p}_{\Lambda})W^2(kR_{8,\Lambda})\right]^{-1}\,,
\end{small}
\end{equation}
where the chosen values of the parameters in this expression define a fiducial model. Specifically, we have set $\delta_{m,\Lambda}\equiv\delta_{m,\Lambda}(z=0)$ and the parameters of the vector $\vec{p}_\Lambda$ are taken to be equal to those from the Planck 2015 TT,TE,EE+lowP+lensing analysis (Planck Collab. XIII 2016). The subscript $\CC$ in all these parameters denotes such a setting. In particular,  $\sigma_{8,\Lambda}$ in \eqref{eq:P0} is also taken from the aforementioned Planck 2015 data.  However, $\delta_{m,\Lambda}$ in the same formula is computable: it is the value of $\delta_m(z=0)$ obtained from solving the perturbations equation of the $\CC$CDM, i.e. Eq. \eqref{eq:DensityContrastEq} with $\psi=0$, using the mentioned fiducial values of the other parameters.

Another way of dealing with the normalization of the power spectrum would consist in leaving $P_0$ free in the fitting analysis, while forcing it to satisfy the Planck 2015 CMB bounds. The point is that the Planck Collaboration does not provide such constraints. Alternatively, they provide the central value and associated uncertainty of the $A_s$ parameter, i.e. the normalization factor of the (dimensionless) primordial power spectrum of the scalar perturbations,
\begin{equation}
\mathcal{P}_{\mathcal{R}}(k)\equiv A_s\left(\frac{k}{k_*}\right)^{n_s-1}\,,
\end{equation}
where $k_*=0.05\,{\rm Mpc}^{-1}$ is Planck's pivot scale (Planck Collab. XIII 2016). The relation between  $A_s$ with $P_0$   can be found using standard formulae (see e.g. Gorbunov \& Rubakov 2011; Amendola \& Tsujikawa 2015). We find:
\begin{equation}
P_0=A_s\frac{8\pi^2}{25}\frac{k_*^{1-n_s}}{(\Omega_m h^2)^2(100\varsigma)^4}\,,
\end{equation}
with $\varsigma\equiv 1$ km/s/Mpc$=2.1332\times10^{-44} GeV$ (in natural units), and  $h$ is defined as usual through $H_0=100 h\, \varsigma$.  Let us note that both $P_0$ and $A_s$  encode information of the primordial universe, whereas $\sigma_8$ strongly depends on the physics of the late-time expansion and, therefore, on the features of the DE or vacuum energy, in particular of its possible time-evolution. It is thus natural to rely on the Planck 2015 constraint on $A_s$ rather than on $\sigma_8$, since the latter is clearly more sensitive to the $\Lambda$CDM assumption used in Planck's analysis. This is the reasoning that has motivated the fitting scheme followed by us, in which $\sigma_8$ is a computed quantity from the fitting parameters of Table 1 and Eq.\,(\ref{eq:s88generalNN})  rather than picking up some a priori fiducial value. Notice, also, that the constraint extracted from the Planck 2015 TT,TE,EE+lowP+lensing analysis is $10^9 A_s=(2.130\pm 0.053)$. It is worth remarking that the uncertainty on the value of $A_s$ is only of $\sim 2.5\%$. Thus, a variation of $A_s$ respecting the tight margin left by that constraint is completely unable to account for the observed deficit of structure formation in the context of the $\Lambda$CDM, as we have checked. In other words, this narrow freedom cannot be used to relax the $\sigma_8$-tension in the context of the $\CC$CDM  (cf. Fig. 4, where we show that the needed relative change in $f(z)\sigma_8(z)$ is around  $\sim 8\%$). Lower values of $A_s$ (or, equivalently, of $P_0$) would be of course very welcome by the LSS data, since the theoretical curve of $f(z)\sigma_8(z)$ would be lowered, but such values value of the power spectrum normalization would then be tensioned with the Planck 2015 CMB constraint. We conclude that a variation of $A_s$ is unable to explain alone,  in a consistent way, the needed reduction in $f(z)\sigma_8(z)$.  Such cul-de-sac situation for the $\CC$CDM  suggests that a new dynamical variable beyond the $\CC$CDM may be necessary to account for the $\sigma_8$-tension. We propose that the needed variable is connected with the dynamical character of the DE, in contrast to the rigid status of $\CC$ in the $\CC$CDM.  In the next section we illustrate the  benefits that are obtained concerning the  $\sigma_8$-tension if we adopt a DDE point of view.   In the previous sections we have prepared the ground for such calculation both at the background and perturbations level. By performing a detailed computation of the growth rate within the RVM we find that   $\sigma_8$ becomes reduced by precisely the desired amount of $\sim 8\%$, if we use the fitting values from Table 1. We also compare with the corresponding result within the XCDM.

\section{Analytical calculation of  $\Delta_{\lowercase{f}\sigma_8}(\lowercase{z})$ in the RVM: solving the $\sigma_8$-tension}

In this section we provide a detailed analytical calculation, supported by numerical analysis, aimed at explaining  how and why the RVM is capable to produce the necessary  $\sim 8\%$ reduction of $\sigma_8$, and in general of $f(z)\sigma_8(z)$, with respect to the $\CC$CDM. It is well-known that the  $\CC$CDM predicts a too large value of $\sigma_8$ and hence an exceeding structure formation power that is unable to explain the LSS data represented by the  $f(z)\sigma_8(z)$ observations, see our Fig. 4.  In the following we will show how the vacuum coefficient $\nu$ of the RVM is capable to provide the necessary  $\sim 8\%$  decrease despite its fitted value (cf. Table 1) is of order $\nu\sim 10^{-3}$.  Let us also mention at this point that there are a few alternative approaches attempting to cure the $\sigma_8$-tension, e.g. using possible effects of viscosity of the cosmic fluid (Anand et al. 2017), or some phenomenological interactions between DM and DE (Barros et al. 2018; An,  Feng \& Wang  2017; Wang et al. 2016), or even using a small amount of spatial curvature (Ooba, Ratra \& Sugiyama 2017).  Another potentially significant effect comes from the impact of massive neutrinos, see the studies by Hamann \& Hasenkamp 2013;  Battye \&  Moss 2014: Salvatelli et al. 2014, and the recent works by Lorenz, Calabrese \& Alonso 2017  and Mishra-Sharma, Alonso \& Dunkley 2018. In fact, dynamical dark energy models may exhibit degeneracies with the cosmic neutrino background since massive neutrinos can suppress the power spectrum (and hence the structure formation) on small scales (see Hu, Eisenstein \& Tegmark 1998; Shoji \& Komatsu 2010). However, the above mentioned papers show that the effect proves insufficient to relax the $\sigma_8$-tension, if the allowed neutrino mass hierarchies are to be respected.  Other recent works have examined this problem within particular DDE models (e.g. Guo, Zhang \& Zhang 2018; Park \& Ratra 2018; McCarthy et al. 2018).  In another vein, it has been suggested that one can mitigate the tension by allowing the amplitude of the CMB lensing power spectrum, $A_{Lens}$, to be free when fitting the TT power spectrum rather than fixing its natural value to unity, what might reflect an unaccounted for systematic issue, see e.g. Addison et al. 2017 and McCarthy et al. 2018. These various possibilities deserve of course further examination, but here we wish to put the emphasis on the impact from dynamical vacuum energy and in particular within the framework of the RVM. It turns out that a detailed study of this problem within the RVM  is feasible both at the analytical and numerical level and we shall show next that the results are perfectly consistent. The remarkable outcome is that vacuum dynamics alone can dispose of the $\sigma_8$-tension.  Subsequent studies on the interplay between the various types of mentioned alternative effects  should  be interesting, of course, but they are beyond the reach of the current study. 

The solution to the $\sigma_8$-tension that we are proposing  here with the help of the RVM  was first advanced by us in (G\'omez-Valent \& Sol\`a 2018). It is truly an economical and efficient solution, in the sense that the tension becomes fully relaxed.  This result is not obtained by just focusing exclusively on the LSS data but by considering a global  quality fit to the entire string of SNIa+BAO+$H(z)$+LSS+CMB  observations. The fit quality of the RVM is substantially better than that of the $\CC$CDM, see Table 1.

\begin{figure*}
\includegraphics[scale=0.85]{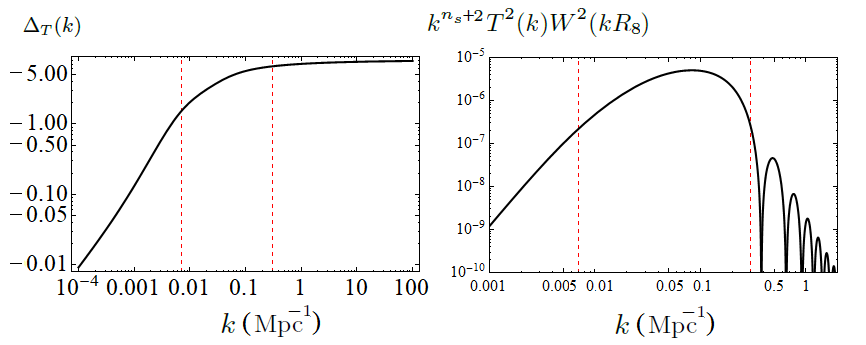}
\caption{{\it Left plot:} Relative difference in the transfer function \eqref{eq:BBKS} between the RVM (cf. Table 1) and the $\Lambda$CDM (with $\nu=0$ but the other parameters chosen equal to the RVM ones) as a function of $k$, i.e. $\Delta_T(k)=100\cdot (T_{\rm RVM}(k)-T_\Lambda(k))/T_{\Lambda}(k)$; {\it Right plot:} Product of functions entering the integral of \eqref{eq:s88generalNN}, also as a function of $k$. The range of wave numbers at which this product gets its largest values is the most sensitive range to the relative differences induced by a $\nu$ in the transfer function. We have marked off this approximate range of $k$'s by red vertical dashed lines in both plots in order to ease the visualization.
\label{fig:RVMdifOrigin4}}
\end{figure*}

First of all let us focus our attention on the left plot of Fig. 5. It is aimed to show the individual impact of each parameter on the $f(z)\sigma_8(z)$ observable. A short description of this plot is in order. We take as baseline model the $\Lambda$CDM with the fitted values provided in Table 1. The corresponding curve is the black one. To obtain the dotted blue curve, we have only changed the value of $n_s$ with respect to the reference line, and have set it to the best-fit value of the RVM. The curve moves mildly upwards.  We see that the effect of this change is derisory ($\sim 1\%$). The other curves plotted therein are obtained upon progressively setting the various parameters to the values obtained in the fitting
analysis of the RVM (cf. Table 1).  If we do not only change the value of $n_s$, but also set $\Omega_m$ to the RVM value, we obtain the dashed red curve (labelled $n_s+\Omega_m$), which is noticeably higher. Clearly these two changes push the prediction in the wrong direction since the resulting curves are shifted upwards and therefore imply even higher structure formation power than  the concordance model (the black curve). As indicated, the remaining curves are obtained by sequentially incorporating the changes in the other parameters to the previous configurations, analogously to the procedure described before. The next change is going to revert the ``wrong'' movements made before. Indeed, the dashed purple curve (referred to as $n_s+\Omega_m+h$ in the legend), which is obtained upon adding the change in $h$ to the previous situation, lies now very near (just slightly below) to the $\Lambda$CDM one (the relative difference is only about $\sim 2\%$). This means that the change in $h$ is significant enough as to counteract the previous unfavorable changes. Later on we will show how this comes about. Concerning $\omega_b$, its variation has  almost no effect on $f(z)\sigma_8(z)$, and the corresponding curve (labelled $n_s+\Omega_m+h+\omega_b$ and in orange) lies just on top of the last one. What finally makes a big difference to bring the theoretical curve towards the correct direction is the role played by the  vacuum parameter $\nu$. This can be easily appraised by direct comparison of the orange curve and the brown continuous curve (the latter contains, in addition to the former, the effect of $\nu$). In point of fact, a non-null and positive value of $\nu$ is the genuine force capable of dragging the theoretical  $f(z)\sigma_8(z)$  curve downwards as a whole by the desired  amount ($\sim 8\%$ ) so as to conform with the data points, and therefore we can assert that $\nu$ is the crucial new ingredient that warrants a fit to the LSS data better than the $\CC$CDM. Of course, $\nu$ can depart from zero because the other parameters can be readjusted without worsening the fit to the other data sets, and this is also very important. In particular, let us recall that the product $\omega_m\equiv\Omega_m h^2$ is very much constrained by the CMB data, and this fact enforces $\omega_m$ to remain very near the value 0.141.  In our case we find $\omega_m=0.1412$ for the $\CC$CDM. Actually, we find that the relation $\omega_m=0.141$ defines a degeneracy curve in the $\Omega_m-h$ plane, meaning that if we move on this curve by varying $\Omega_m$ (or $h$) at fixed $\omega_m=0.141$, we find no significant changes in the prediction of $f(z)\sigma_8(z)$  for the $\Lambda$CDM. This is shown in the right plot of Fig. 5 where all curves crowd around the black one.

It is important to understand that the possible modifications of the weighted growth rate caused by a dynamical vacuum scenario can be  important only in the recent universe, where the DE starts to dominate over the CDM. The mentioned degeneracy is only approximate near the present time, of course, but it helps to understand how $\Omega_m$ and $h$ can both vary while respecting the CMB bounds and at the same time keeping almost intact the $f(z)\sigma_8(z)$ curve predicted by the concordance model. The RVM, however, can break this degeneracy thanks to the vacuum parameter $\nu$,  which for small positive values can bring the LSS curve down, relaxing in this way the well-known tension between the $\Lambda$CDM and the LSS data.

\begin{figure*}
\includegraphics[scale=0.7]{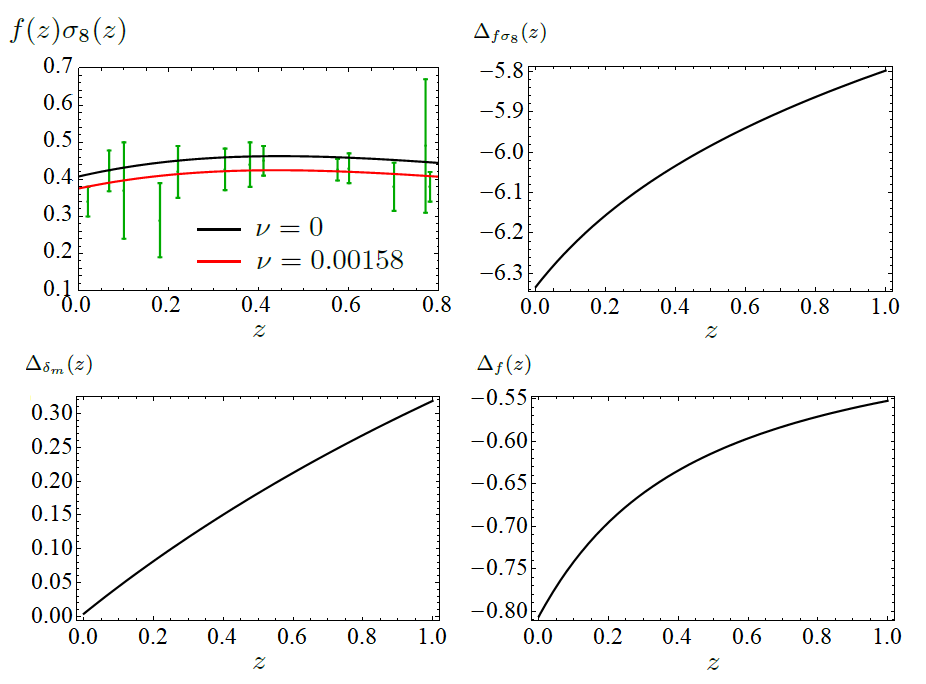}
\caption{{\it Upper-left plot:} Weighted growth rate obtained by (i) setting all the parameters to the RVM ones (cf. Table 1); and (ii) keeping the same configuration, but with $\nu=0$. These correspond to the red and black lines, respectively; {\it Upper-right plot:} Relative difference between the curves of the upper-left plot, as defined in \eqref{eq:Deltafsigma8}. The change induced by the non-null vacuum parameter reaches  $\sim 6.3\%$ at $z\sim 0$; {\it Lower-left plot:} Relative difference between the density contrasts $\delta_m(z)$ associated to the two scenarios explored in the upper-left plot, expressed in $\%$. The differences in this case are lower than $0.4\%$ for $z<1$; {\it Lower-right plot:} The same, but for the growth function $f(z)$, in $\%$ too. Around the present time, the relative differences attain the $0.8\%$ level.
\label{fig:RVMdifOrigin4}}
\end{figure*}
%

Now that we have demonstrated graphically the crucial role played by $\nu$ in the needed lowering of the theoretical $f(z)\sigma_8(z)$ curve, we can proceed to study which is the analytical explanation for the fact that a tiny parameter of order $10^{-3}$ can induce changes in the LSS prediction one order of magnitude larger than $\nu$ itself. To this end let us start by computing the leading order corrections induced by $\nu$ in the matter transfer function \eqref{eq:BBKS}. The percentage change caused by a non-null $\nu$ (if we keep the other parameters constant) can be computed as follows:
\begin{equation}\label{eq:DeltaTdef}
\Delta_T(k,\nu)=\frac{100}{T_\Lambda}\frac{\partial T(x)}{\partial\nu}\bigg\rvert_{\nu=0}\nu+\mathcal{O}(\nu^2)\,,
\end{equation}
where $T_{\Lambda}\equiv T(x_\CC)$, $x_\CC\equiv k/k^\CC_{eq}$. Let us firstly compute the correction $\Delta_T$ when $x\gg 1$ or, equivalently, when $k\gg k_{eq}$, and see whether we can extract information from this calculation. In this limit the BBKS transfer function \eqref{eq:BBKS} can be approximated just by
\begin{equation}
T(x)\approx \frac{C\ln(1+Ax)}{x^2}\,,
\end{equation}
with $A=0.171$ and $C=(0.171\times 0.49)^{-1}$. In order to compute \eqref{eq:DeltaTdef} it is convenient to use the differentiation chain rule:
\begin{equation}\label{eq:3terms}
\frac{\partial T(x)}{\partial\nu}\bigg\rvert_{\nu=0}=\frac{\partial T(x)}{\partial x}\frac{\partial x}{\partial k_{eq}}\frac{\partial k_{eq}}{\partial\nu}\bigg\rvert_{\nu=0}\,.
\end{equation}
The first factor on the ${\it r.h.s.}$ of this relation reads,
\begin{equation}
\frac{\partial T(x)}{\partial x}\bigg\rvert_{\nu=0}=-\frac{2}{x_\Lambda}T(x_\Lambda)+\frac{CA}{x_\Lambda^2(1+Ax_\Lambda)}\,.
\end{equation}
Taking into account that $k^\CC_{eq}\sim 0.01\,{\rm Mpc}^{-1}$, it is easy to see that for $k\gtrsim 1\,{\rm Mpc}^{-1}$ ($x\gtrsim 100$) the second term in the last expression can be neglected and therefore:
\begin{equation}\label{eq:T1}
\frac{\partial T(x)}{\partial x}\bigg\rvert_{\nu=0}\approx-\frac{2}{x_\Lambda}T_\Lambda\,.
\end{equation}
The second factor on the ${\it r.h.s.}$ of \eqref{eq:3terms} is just
\begin{equation}\label{eq:T2}
\frac{\partial x}{\partial k_{eq}}\bigg\rvert_{\nu=0}=-\frac{x_\Lambda}{k^\CC_{eq}}\,,
\end{equation}
and the last factor  can be obtained upon differentiation of \eqref{eq:keqRVM} with respect to $\nu$,
\begin{equation}\label{eq:T3}
\frac{\partial k_{eq}}{\partial\nu}\bigg\rvert_{\nu=0}=k^\CC_{eq}\left[-\frac{7}{2}+3\ln\left(\frac{\Omega_r}{\Omega_m}\right)\right]\,.
\end{equation}
Introducing \eqref{eq:T1}, \eqref{eq:T2} and \eqref{eq:T3} in \eqref{eq:3terms} we finally obtain:
\begin{equation}\label{eq:DeltaT}
\Delta_T (x\gg 1) =- 100\nu\left[7+6\ln\left(\frac{\Omega_m}{\Omega_r}\right)\right]+\mathcal{O}(\nu^2)\,.
\end{equation}
By using in the above formula the values of the RVM parameters presented in Table 1, we see that the asymptotic relative difference between the RVM and the $\Lambda$CDM transfer functions is constant and attains  $-8.8\%$. This is precisely the number that we get for the asymptotic value of $\Delta_T$ in our numerical results (cf. the left plot of Fig. 6). Of course, \eqref{eq:DeltaT} might still not allow us to directly infer the ultimate correction induced by $\nu$ on the value of $\sigma_8$ since there are other contributions to consider. To ease the discussion let us write symbolically Eq.\,(\ref{eq:s88generalNN}) in the form $\sigma_8=\delta_m\sqrt{I}$, where $I$ is the integral over  the wave number $k$ involved in that equation.  Obviously,  the transfer function is only part of the integrand of $I$,  and to assess the relative correction on $\sigma_8$  we have to evaluate the relative corrections  $\Delta_{\delta_m}$ and  $\Delta_{\sqrt{I}}$  on  each one of the factors.  Let us therefore study this issue more carefully. In the right plot of Fig. 6 we show  the shape of the relevant function in the integrand of $I$, i.e.  $k^{n_s+2} T^{2}(k) W^2(kR_{8})$.  It is clear from the plot that for wave numbers $k\gtrsim 0.5\,{\rm Mpc}^{-1}$ the integrand is very suppressed, whereas it is much more sizeable for a range of smaller wave numbers where it reaches a maximum, specifically in the range $0.007\,{\rm Mpc}^{-1}\lesssim k \lesssim 0.3\,{\rm Mpc}^{-1}$. This range has been marked off with red vertical dashed lines in both plots of Fig. 6 to facilitate the reading of the results. According to the left plot of Fig.\,6 and the aforesaid range of relevant wave numbers, we expect $\Delta_{\sqrt{I}}$ to be around $-5.5\%$.  Next we have to sum to it the  numerical contribution  from the density contrast, i.e. $\Delta_{\delta_m}$ (c.f. the lower-left plot of Fig. 7),  so as to obtain the net effect  $\Delta_{\sigma_8}$. Finally, the total relative correction undergone by $f(z)\sigma_8(z)$ induced by the presence of the vacuum parameter $\nu$  is given by
\begin{equation}\label{eq:Deltafsigma8Total}
\begin{array}{ll}
\Delta_{f\sigma_8}(z)& = \Delta_f(z)+\Delta_{\sigma_8}(z) \\
   & =\Delta_f(z)+\Delta_{\delta_m}(z)+\Delta_{\sqrt{I}}\,.
\end{array}
\end{equation}
where $\Delta_f$ (computed numerically in the lower-right plot of Fig. 7) is the corresponding contribution from the growth rate. The total percentage correction (\ref{eq:Deltafsigma8Total}) is shown in the upper-right plot of the same figure.
%
\begin{figure}
\includegraphics[scale=0.65]{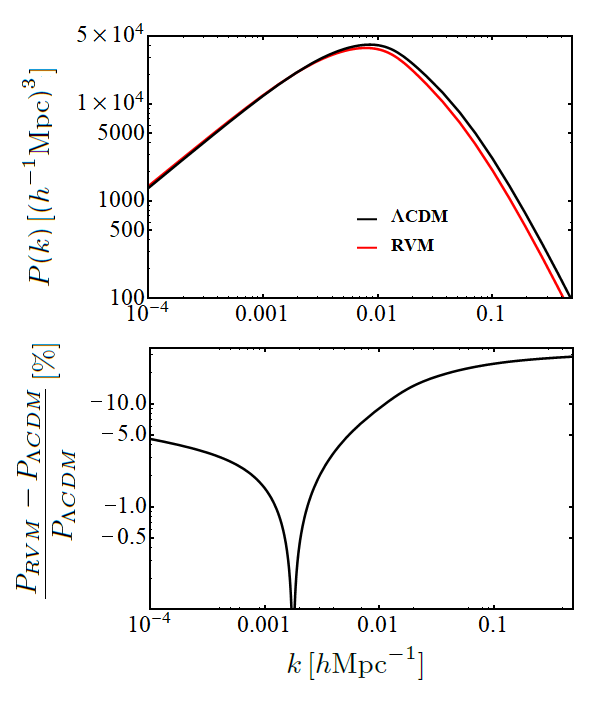}
\caption{{\it Upper plot:} The present linear power spectrum of matter perturbations $P(k)$ obtained for the $\Lambda$CDM and the RVM model; {\it Lower plot:} The relative difference between the two power spectra expressed in \%.
\label{fig:P(k)}}
\end{figure}
%
%
Using this formula, together with our theoretical estimation $\Delta_{\sqrt{I}}\approx -5.5\%$, which is redshift independent, and the numerical results for $\Delta_{\delta_m}(z)$ and $\Delta_f(z)$ shown in the lower plots of Fig. 7, we can check that we recover the total relative difference $\Delta_{f\sigma_8}(z)$ shown in the upper-right plot of the same figure. For instance, for $z=0$ we find:
\begin{equation}
\Delta_{f\sigma_8}(0)\approx -0.8\%+0\%-5.5\%=-6.3\%\,,
\end{equation}
and for $z=0.8$:
\begin{equation}
\Delta_{f\sigma_8}(0.8)\approx -0.55\%+0.3\%-5.5\%=-5.75\%\,,
\end{equation}
which match almost perfectly with the values of the upper-right plot of Fig. 7. Let us note that the obtained result  is not the real correction predicted by the RVM, which is around $-8\%$ (cf. Fig. 4 right) because the parameters of the $\CC$CDM are also fixed at the central values of the RVM, but with vanishing $\nu$.

Therefore we can say that we have been able to identify the origin of the lowering of the $f(z)\sigma_8(z)$ curve in the RVM. The most part of the effect ($\sim 75\%$) is driven by $\nu$. More concretely, $\sim 65\%$ of the induced changes are due to the negative shift in the value of $k_{eq}$, which directly translates into a negative shift of the location of the power spectrum's maximum (cf. formula \eqref{eq:keqRVM} and Fig. 8), see also the analysis of (Perico \& Tamayo 2017) and (Geng, Lee \& Yin 2017). Although the effect of $\nu$ in the evolution of the density contrast at subhorizon scales is not exaggeratedly big (the corrections are of order $\nu$, as expected), the RVM parameter $\nu$  is capable of changing in a significant way the time at which the different modes reentered the Hubble horizon with respect to the concordance case.  Despite $\nu$ is small, the scale factor and wave number at the equality point between matter and radiation epochs become modified in a non-negligible way, see Eqs.\,(\ref{eq:aeqRVM}) and (\ref{eq:keqRVM}).  As a consequence, the LSS formation is substantially suppressed at low scales with respect to the $\Lambda$CDM and in this way the tension with the $f(z)\sigma_8(z)$ data loosens, mainly due to an important decrease of the $\sigma_8$ parameter. Let us stress that such feature cannot be appraised if one restricts the analysis mostly to the CMB without sufficient LSS input (Heavens et al. 2017; Perico \& Tamayo 2017).
%

Contrary to the RVM, the XCDM can only lower $f(z)\sigma_8(z)$ roughly by $2-4\%$ at most with respect to the $\Lambda$CDM (cf. Fig. 4). In this parametrization  we have $\rho_X(a)=\rho_{X0}a^{-3(1+w)}$, with $\rho_{X0}=\rLo$ and the EoS parameter satisfying  $w\ne -1$.
We can perfectly explain why the XCDM parametrization cannot match the very good description of the LSS data by the RVM. The reason is pretty simple. A DE parameter $w$ close (but not equal to) $-1$ cannot change the transfer function, just because the equality time between matter and radiation energy densities is not modified (matter and radiation are covariantly self-conserved) and the contribution of DE to the critical energy density at $a_{eq}$ is completely negligible. Thus $k_{eq}$ is not sensitive to $w$ and $T(k)$ remains unaltered. This leads us to conclude that the only modifications induced by $w$ on the $f(z)\sigma_8(z)$ observable can be due to late-time physics, mainly through the changes in the density contrast and the growth rate caused by the late-time domination of the DE over the non-relativistic matter. These corrections are of a few percent and cannot give rise to the desired level of lowering of the $f(z)\sigma_8(z)$ curve. By taking a look on the reconstruction plot of Fig. 9, which is the analogous of the left plot of Fig. 5, we  can observe that for the XCDM the final effect is mainly due to the deviation of $w$ from $-1$. We also find that the compensation between $\Omega_m$ and $h$ discussed before for the RVM also occurs here.

\section{Dynamical dark energy: LSS data versus weak-lensing data}

The improvement in the description of the LSS data in the XCDM, and more conspicuously in the RVM, does not only concern the $f(z)\sigma_8(z)$ data, but also some weak gravitational lensing constraints on the conventional quantity  $S_8\equiv \sigma_8(\Omega_m/0.3)^{0.5}$ that one can find in the literature (see e.g. Heymans et al. 2013; Hildebrandt et al. 2017; Joudaki et al. 2018). The impact of the DDE is crystal-clear from the left plot of Fig. 10, where we show the contour lines in the ($\Omega_m,\sigma_8)$ plane obtained from the very same datasets used in the fitting analyses presented in Table 1 for the $\CC$CDM, the XCDM and the RVM, together with the observational constraints in the same plane provided by: (i) DES Collab. 2017, extracted from weak gravitational lensing tomography, $S_8=0.783^{+0.021}_{-0.025}$; (ii)  Joudaki et al. 2018, $S_8=0.742\pm 0.035$, obtained by KiDS-450, 2dFLenS and BOSS collaborations from a joint analysis of weak gravitational lensing tomography and overlapping redshift-space galaxy clustering; and (iii) KiDS-450 collaboration (K\"ohlinberg et al. 2017), obtained from weak gravitational lensing tomography, $S_8=0.651\pm 0.058$. The last two data points on $S_8$ tend to favor lower values of $\sigma_8$. Very similar results have been found using only weak gravitational lensing tomography data by KiDS-450 collaboration (Hildebrandt et al. 2017), $S_8=0.745\pm 0.039$, and also by CFHTLenS (Heymans et al. 2013), $(\Omega_m/0.27)^{0.46}=0.770\pm 0.040$. In contrast, the point provided by DES is more resonant with Planck, but due to its large uncertainty it is still fully compatible with  Joudaki et al. 2018; Hildebrandt et al. 2017; and Heymans et al. 2013. Our discussion on the ability of the models under study to describe the gravitational weak lensing data basically remains unchanged if we use the constraints from Heymans et al. 2013 or Hildebrandt et al. 2017 instead that of Joudaki et al. 2018.

\begin{figure}
\includegraphics[scale=0.45]{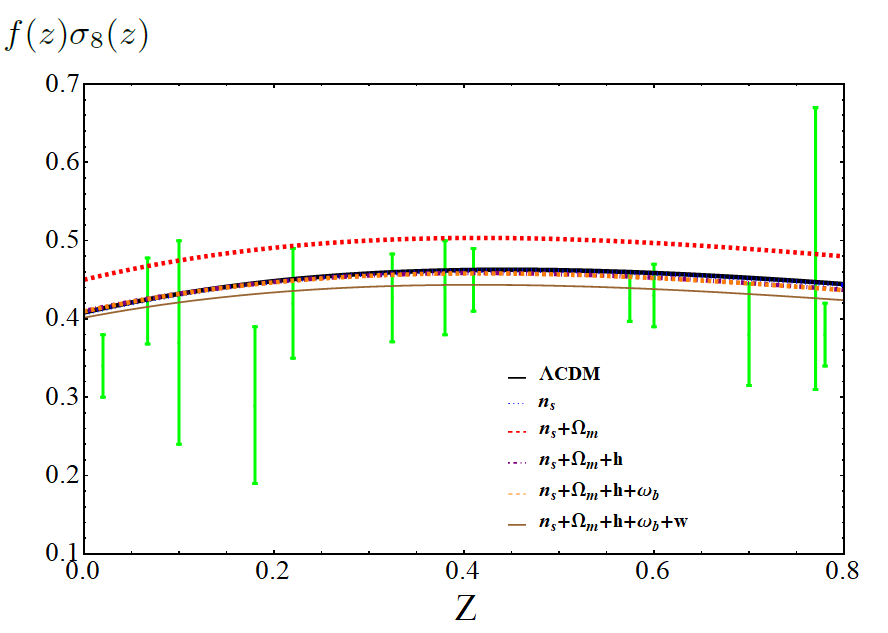}
\caption{Reconstruction of the $f(z)\sigma_8(z)$ curve of the XCDM from the $\Lambda$CDM one, following the same procedure utilized in Fig. 5 (left).
\label{fig:XCDMcons}}
\end{figure}
In Fig. 10 we can better assess the impact of the weak-lensing data. Two comments related with the results  shown in that figure are in order: (i) the $\Lambda$CDM is compatible with the $S_8$ data point of  Joudaki et al. 2018 at  $1\sigma$ only, so the tension of the $\CC$CDM with $S_8$ is actually very small. Despite this, it is intriguing to observe that the RVM achieves an outstanding level of concordance with this data point. It actually removes completely the existing $1\sigma$ tension between it and the concordance model. If that is not enough, the RVM best-fit value is almost centered in the band delimited by the dashed curves in purple that corresponds to the preferred values of (Joudaki et al. 2018); and (ii) the data points from  Heymans et al. 2013; Hildebrandt et al. 2017; and  Joudaki et al. 2018  are also compatible at $1\sigma$ with the constraints obtained by other weak lensing studies such as those by (DES Collab. 2017) or (K\"ohlinger et al. 2017). These have been drawn in green and orange, respectively. The current variety of data points on $S_8$ unavoidably casts a shadow of doubt about the level of confidence that we can ultimately have on the weak lensing constraints in general, since it seems that there exists a non-negligible degree of dispersion of alternative constraints on $S_8$ around the combined value found by  Joudaki et al. 2018 from KiDS-450+2dFLenS+BOSS.  The present situation indicates that the constraints that we can derive  from the LSS data points $f(z)\sigma_8(z)$ seem to be in very good accordance with the weak lensing constraints on  $S_8$  furnished in  the works by Heymans et al. 2013; Hildebrandt et al. 2017; and Joudaki et al. 2018, what is not too surprising since the latter favor lower values of $\sigma_8$ in the range $\sim 0.730-0.750$, rather than the typical values found by the  DES Collab. 2017  (or K\"ohlinger et al. 2017), which tend to favor values of $S_8$ that are larger (respectively, lower) than those inferred from the direct  $f(z)\sigma_8(z)$ data.

\begin{figure*}
\includegraphics[scale=0.6]{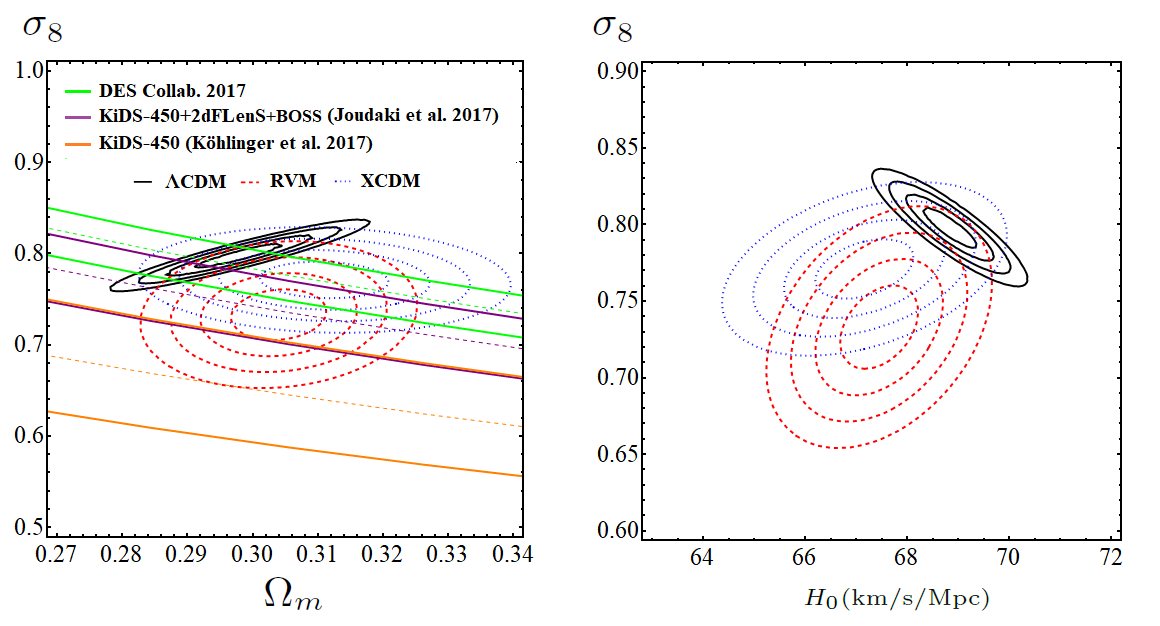}
\caption{{\it Left plot:} Likelihood contour lines in the ($\Omega_m,\sigma_8)$ plane for the values $-2\ln \mathcal{L}/\mathcal{L}_{\rm max}$= $2.30$, $6.18$, $11.81$, and $19.33$ (corresponding to $1\sigma$, $2\sigma$, $3\sigma$, and $4\sigma$ c.l.) obtained from the very same fitting analyses presented in Table 1 for the $\Lambda$CDM, the XCDM and the RVM, together with the observational constraints in the same plane provided by: (i) (DES Collab. 2017), extracted from weak gravitational lensing tomography, $S_8=0.783^{+0.021}_{-0.025}$ (green curves); (ii) combined values of KiDS-450+2dFLenS+BOSS collaborations, extracted by Joudaki et al. 2018 from weak gravitational lensing tomography and overlapping redshift-space galaxy clustering, $S_8=0.742\pm 0.035$ (purple curves); (iii) KiDS-450 collaboration (K\"ohlinberg et al. 2017), obtained from weak gravitational lensing tomography, $S_8=0.651\pm 0.058$ (orange curves). We show the allowed $1\sigma$ bands for the three data points on $S_8$ used; {\it Right plot:} Contour lines up to $4\sigma$ c.l. in the ($H_0,\sigma_8)$ plane for the same three models under study.
\label{fig:CLcombi}}
\end{figure*}

It may be appropriate to single out at this point the recent and interesting works by Lin  \&  Ishak 2017a,b, in which the authors run a so-called (dis)cordance test based on using a proposed index of inconsistency (IOI)  tailored at finding possible inconsistencies/tensions between two or more data sets in a systematic and efficient way. For instance, it is well-known that there is a persistent discrepancy between the Planck  CMB measurements of $H_0$ and the local measurements based on distance ladder (Riess et al. 2016, 2018b). At the same time, if one compares what is inferred from Planck 2015 best-fit values, the LSS/RSD measurements generally assign smaller power to the large scale structure data parametrized in terms of the weighted linear growth rate $f(z)\sigma_8(z)$.  This feature is of course nothing but the $\sigma_8$-tension we have been addressing in this paper.  It is therefore natural to run the IOI test for the different kinds of $H_0$ measurements and also to study the consistency between the $H_0$ and the growth data.   For example, upon comparing the constraints on $H_0$  from different methods Lin  \&  Ishak 2017b  observe a decrease of the IOI when the local $H_0$-measurement is removed. From this fact they conclude that the  local measurement of $H_0$  is an outlier compared to the others, what would favor a systematics-based explanation. This situation is compatible with the observed improvement in the statistical quality of the fitting analysis by Sol\`a, G\'omez-Valent \& de Cruz P\'erez 2017c,d  when the local $H_0$-measurement is removed from the overall fit of the data using the RVM and the $\CC$CDM.  In this respect, let us mention that a recent model-independent analysis of data on cosmic chronometers and an updated compilation of SNIa seem to favor the lower range of $H_0$ (G\'omez-Valent \& Amendola 2018), what would be more along the line of the results found here, which favor a theoretical interpretation of the observed $\sigma_8$ and $H_0$  tensions  in  terms of vacuum dynamics and in general of DDE (cf. Fig. 10).

The mentioned authors of the IOI test  actually apply it to two large sets of current observational data:  the geometry data  (e.g. SNIa, BAO etc.) versus the growth data (e.g. LSS/RSD, weak-lensing, CMB-lensing etc.). They find that a persistent inconsistency is present between the two sorts of data sets. Despite encountering such inconsistency, Lin  \&  Ishak 2017a,b emphasize that if they focus on the LSS data sets (which include e.g. the WiggleZ power spectrum, SDSS redshift space distortion, CFHTLenS weak lensing, CMB lensing, and cluster count from SZ effect) there is a global consistency among them. They confirm they are consistent one with another and also when all combined. In contrast, they find a persistent moderate inconsistency between Planck and individual or combined LSS probes.  For the time being this cannot be fixed within the $\CC$CDM.  However,  if we  combine the fact that the RVM fit of  $H_0$ (Sol\`a, G\'omez-Valent \& de Cruz P\'erez 2017c,d)  is in good accordance with the Planck value, and the result obtained here showing that the RVM is also capable of relaxing the $\sigma_8$-tension (in contrast to the $\CC$CDM), it seems judicious to conclude that the running vacuum model can furnish a successful joint description of the observables $H_0$ and $\sigma_8$.  For this reason when we consider the various sources of weak-lensing data discussed above we tend to prefer those that are more in accordance with the direct LSS growth data, such as e.g. the weak-lensing data from  Joudaki et al. 2018,  rather than the weak-lensing data from DES Collab. 2017, which are more in accordance with the (tensioned) $\sigma_8$-values predicted by Planck.

To summarize, despite that gravitational lensing statistics has since long been considered  as a possible probe for the EoS of the DE (Cooray \& Huterer  1999) and hence as a useful test for DDE, the bare truth is that at present the wealth of growth data collected from the direct $f(z)\sigma_8(z)$ measurements at different redshifts seem to encode much more accurate information on the possible dynamical nature of the DE.  While the lensing data are  compatible with the growth data, the dispersion of the lensing measurements  is too large at present to provide a firm handle to possible DDE observations.  Thus, in current practice a putative DDE signal from these sources becomes considerably blurred, what is in stark contrast  with the situation involving direct growth data points. In fact, with the help of these data (and the remaining observational sources) we have shown here that even a simple XCDM parametrization enables us to extract a DDE signal at near $3\sigma$ c.l. Interestingly  the signal can be further enhanced within the RVM up to $3.8\sigma$.  At the end of the day the possibility of having running vacuum proves  particularly sensitive to the features of the growth data and  this fact produces a remarkable improvement of the overall fit quality of the RVM as compared to the $\CC$CDM.


\section{Conclusions}

It is well known that the $\CC$CDM harbors important theoretical conundrums, but  is also plagued with some persistent phenomenological problems of very practical nature. To put it in a nutshell, we can say that there is  a significant tension between the geometry data and the growth data. Two representative observables illustrating this tension are the disparate results for $H_0$ obtained from local and CMB measurements, and the exceeding power associated to large scale structure (LSS) formation data, which lead to  the $\sigma_8$-tension. These problems cannot be currently cured within the concordance model with rigid cosmological term $\CC=$const. In this work we have considered a possible way to relax these tensions by admitting the possibility of dynamical dark energy (DDE) models. Most particularly we have focused  on the running vacuum model (RVM), although we have studied the problem also within the simple XCDM parametrization of the DDE. In order to tackle these tensions in a consistent way we have first of all undertaken a careful study of the matter density perturbations in the context of dynamical vacuum models (DVMs). In these models the vacuum fluctuations have to be considered as well, in principle, but we have explicitly shown that they are negligible at all  subhorizon scales that are relevant for the study of the large scale structure formation data. We have considered possible  issues of gauge dependence and we have presented the results both in the conformal Newtonian gauge and in the synchronous gauge. The outcome is that the effective matter perturbations equation obeyed by the density contrast for the DVMs is the same in both gauges and is free both from scale-dependence and from significant vacuum fluctuation effects. This result is valid at scales below the horizon and is similar to the $\CC$CDM case.  The effective equation for DVMs, however, is different from the standard one in the $\CC$CDM and reduces to it in the limit of constant vacuum energy density.

Armed with the previous theoretical results we have faced the practical study and possible resolution of the mentioned tensions between theory and observation in the context of the RVM and compared with the XCDM. While in previous works we had addressed the $H_0$  tension between the Planck CMB data and the local measurements  (Sol\`a, G\'omez-Valent \& de Cruz P\'erez 2017d),  here we have concentrated on  the growth data, namely the observations that are obtained by means of the direct measurements of the weighted growth rate $f(z)\sigma_8(z)$. We have noted that  many studies essentially incorporate the LSS observations only through the gravitational weak-lensing data parametrized in terms of $S_8$ and we have signaled that this practice may result in an insufficient account of the LSS data. We find that the pictures achieved in terms of the  $f(z)\sigma_8(z)$ data and the constraints on $S_8$ from weak-lensing data  point consistently towards the same direction, but are not equivalent. The current $f(z)\sigma_8(z)$ data turn out to be more restrictive than the $S_8$ data  insofar as concerns the monitoring of a possible DDE signal in the observations. In addition, the latter are not able to improve in a significant way the constraints on the cosmological parameters which we had previously obtained from a rich string of SNIa+$H(z)$+CMB+BAO+$f(z)\sigma_8(z)$ observables  (Sol\`a, G\'omez-Valent \& de Cruz P\'erez 2017d), as we have explicitly checked here.   At the end of the day the weak lensing data can be regarded as a useful complementary source of LSS information, which we find to be compatible with the $f(z)\sigma_8(z)$ data but the former cannot, at present, be a replacement for the latter. From our study we conclude  that for the time being  only the direct $f(z)\sigma_8(z)$ observations offer the possibility of extracting  the signature of vacuum/DE dynamics when combined with CMB and BAO data, whereas if $S_8$-data  is utilized as a substitute for $f(z)\sigma_8(z)$ in the overall fit it yields a much blurred description of the DDE signal.  In this work we have illustrated the extraction of a possible such signal using the RVM and the XCDM parametrization. For the RVM the $\sigma_8$-tension with the LSS data becomes fully relaxed, whereas  for the XCDM  we observe  a correct trend towards a further relaxation as compared to the $\CC$CDM, but the loosening of the tension is definitely weaker. We interpret these results as new signs of evidence of DDE in modern cosmological observations, along the lines of those that were first reported in the  works by  Sol\`a, G\'omez-Valent \& de Cruz P\'erez 2015, 2017a,b; Sol\`a, de Cruz P\'erez \& G\'omez-Valent 2018; and independently by  Zhao et al. 2017.


\section{Acknowledgements}

We are partially supported by MINECO FPA2016-76005-C2-1-P, Consolider CSD2007-00042,  2017-SGR-929 (Generalitat de Catalunya) and  MDM-2014-0369 (ICCUB). AGV wants to express his gratitude to the Institute of Theoretical Physics of the Ruprecht-Karls University of Heidelberg for the financial support and hospitality during part of the elaboration of this paper.


\appendix

\section{Matter perturbations with baryon conservation}
To derive the equation for the total matter density contrast $\delta_m$ in the presence of a dynamical vacuum component at deep subhorizon scales in the Newtonian gauge, \eqref{eq:DensityContrastEq}, we have solved the system formed by Eqs. \eqref{eq:ExtraRelation}-\eqref{eq:EulerNewton3} and \eqref{eq:PoissonNewton2}-\eqref{eq:ContinuityNewton2}. They are valid when vacuum interacts with matter. In the RVM, though, we have assumed that the vacuum does {\it not} interact with baryons, but only with DM. This fact introduces some changes in our system of perturbed differential equations with respect to the ones we have used before, since the conservation equations can be split now as $\nabla^\mu (T^{dm}_{\mu\nu}+T^{\Lambda}_{\mu\nu})=0$ and $\nabla^\mu T^b_{\mu\nu}=0$. Thus, the perturbed continuity equations read,
\begin{align}
\dot{\delta}_{dm}-\frac{\dot{\rho}_\Lambda}{\rho_{dm}}\delta_{dm}&=k^2v_{dm}\,,\label{eq:A}\\
\dot{\delta}_b&=k^2v_b\,,\label{eq:B}
\end{align}
and the Euler ones,
\begin{align}
\dot{v}_b+\mathcal{H}v_b+\Phi&=0\,,\label{eq:C}\\
\dot{v}_{dm}+\mathcal{H}v_{dm}+\Phi&=\frac{\delta\rho_\Lambda+\dot{\rho}_\Lambda v_{dm}}{\rho_{dm}}\,.\label{eq:D}
\end{align}
In order to preserve the usual Euler equation for DM the {\it r.h.s.} of \eqref{eq:D} must cancel, giving rise to
\begin{align}
\dot{v}_{dm}+\mathcal{H}v_{dm}+\Phi&=0\,,\label{eq:D2}\\
\delta\rho_\Lambda=-\dot{\rho}_\Lambda v_{dm}\,.\label{eq:E}
\end{align}
By subtracting \eqref{eq:D2} from \eqref{eq:C} and solving the resulting differential equation in terms of the scale factor one obtains $v_b-v_{dm}=Ca^{-1}$, a decaying mode, with $C$ being an integration constant. Therefore, for the late-time expansion, one can take $v_b=v_{dm}=v_m$, where in the latter equality use has been made of \eqref{eq:vm}. This allows us to retrieve \eqref{eq:EulerNewton3} from \eqref{eq:C} or \eqref{eq:D2}, and also \eqref{eq:ExtraRelation} from \eqref{eq:E}. The Poisson equation \eqref{eq:PoissonNewton2} still holds, but we must derive the continuity equation for the total matter in the limit $v_b,\,v_{dm}\to v_m$. This can be easily done by computing the derivative of $\delta_m=(\rho_{dm}\delta_{dm}+\rho_b\delta_b)/\rho_m$ with respect to the conformal time, $\dot{\delta}_m$, and using Eqs. \eqref{eq:A}-\eqref{eq:B} together with the background continuity equation \eqref{eq:contBackground}, $\dot{\rho}_\Lambda+\dot{\rho}_{dm}+3\mathcal{H}\rho_{dm}=0$ and $\dot{\rho}_b+3\mathcal{H}\rho_b=0$ (the latter two come from the splitting of the former). The resulting expression is just given by \eqref{eq:ContinuityNewton2}. In the last stages of the Universe expansion, when $v_b\sim v_{dm}\sim v_m$, the perturbed equations that rule the evolution of the total matter density perturbations at scales much lower than the horizon in the case in which vacuum interacts only with DM are the same as the ones found when vacuum interacts also with baryons. The differential equation for the total matter density contrast is in both cases given by \eqref{eq:DensityContrastEq}. Similar considerations can be made in the synchronous gauge, which is dealt with in Sect. 5. For vacuum interacting with DM only, equation \eqref{eq:mathringv} is replaced by one in which $\rho_m\rightarrow \rho_{dm}$ and $v_m\rightarrow v_{dm}$, and we can make a similar choice as in \eqref{eq:E} but now differentiating $\rL$ with respect to the cosmic time rather than conformal time. This gives $\mathring{v}_{dm}+H v_{dm}=0$. On the other hand, if baryons are conserved we have $\mathring{v}_{b}+Hv_{b}=0$ automatically, without any extra condition. The solution of both equations is a decaying mode for the respective velocities. Equivalently, if we subtract both equations we find once more $v_b-v_{dm}=C a^{-1}$ and the two velocities asymptote to the same value $v_m\to 0$. This value can be set exactly equal to zero thanks to the residual gauge freedom of the synchronous gauge (cf. Sect. 5). As a consequence the final perturbations equation (\ref{diffeqD}) is also recovered in this case.

Therefore all the results on matter perturbations at subhorizon scales obtained in the paper remain unaltered when the interaction of vacuum with matter occurs only with dark matter and preserves baryons.

\end{document}